\documentclass[10pt]{article}
\usepackage[dvips]{graphicx}  
\usepackage{color}

\begin{document}

\begin{center}
{\Large\bf  Combining NASA/JPL One-Way Optical-Fiber   Light-Speed  Data with  Spacecraft Earth-Flyby Doppler-Shift Data  to Characterise  3-Space Flow  \rule{0pt}{13pt}}\par

\bigskip
Reginald T. Cahill \\ 
{\small\it  School of Chemistry, Physics and Earth Sciences, Flinders University,
Adelaide 5001, Australia\rule{0pt}{15pt}}\\
\raisebox{+1pt}{\footnotesize E-mail: Reg.Cahill@flinders.edu.au}\par
\bigskip

{\small\parbox{11cm}{%
 We combine data from   two high precision NASA/JPL experiments: (i)  the one-way speed of light experiment using optical fibers:   Krisher T.P., Maleki L., Lutes G.F., Primas L.E., Logan R.T., Anderson J.D. and Will C.M.,  Phys. Rev. D, vol 42, 731-734, 1990,  and (ii)  the spacecraft earth-flyby doppler shift data:   Anderson J.D., Campbell J.K., Ekelund J.E., Ellis J. and Jordan J.F.,  Phys. Rev. Lett.,  vol 100, 091102, 2008, to  give the solar-system galactic 3-space average speed of   486km/s in the direction RA=4.29$^h$, Dec=-75.0$^{\circ}$.  Turbulence effects (gravitational waves) are also evident. Data  also reveals the 30km/s orbital speed of the earth and the sun inflow component at 1AU of 42km/s and also 615km/s near the sun, and for the first time, experimental measurement of the 3-space 11.2km/s inflow of the earth.   The NASA/JPL data is  in remarkable agreement with that determined in other light speed anisotropy experiments, such as Michelson-Morley  (1887), Miller (1933), DeWitte (1991), Torr and Kolen (1981), Cahill (2006), Munera (2007), Cahill and Stokes (2008) and Cahill (2009).  
 \rule[0pt]{0pt}{0pt}}}\medskip
\end{center}

\setcounter{section}{0}
\setcounter{equation}{0}
\setcounter{figure}{0}
\setcounter{table}{0}

\markboth{R.T.  Cahill. Combining  One-Way Optical-Fiber   Light-Speed  Data with  Earth Flyby Doppler Data  to Characterise  3-Space Flow}{\thepage}
\markright{R.T.  Cahill. Combining  One-Way Optical-Fiber   Light-Speed  Data with  Earth Flyby Doppler Data  to Characterise  3-Space Flow}

\tableofcontents

\section{Introduction}
In recent years it has become clear, from numerous experiments and observations, that a dynamical 3-space\footnote{The nomenclature {\it 3-space}   is used to distinguish this dynamical 3-dimensional space from other uses of the word {\it space}.} exists \cite{Book,Review}. This dynamical system gives a deeper  explanation for various observed effects  that, until now, have been  successfully described, but not explained,  by the Special Relativity (SR) and General Relativity (GR) formalisms. However it also offers an explanation for other observed effects not described by SR or GR, such as observed light speed anisotropy, bore hole gravity anomalies, black hole mass spectrum and spiral galaxy rotation curves and an expanding universe without dark matter or dark energy.  Herein yet more experimental data is used to further characterise the dynamical 3-space, resulting in the first direct determination of the inflow effect of the earth on the flowing 3-space. The 3-space flow is in the main  determined by the  Milky Way and local galactic cluster.  There are also components related to the orbital motion of the earth and to the effect of the sun, which have already been extracted from experimental data \cite{Book}.

The postulate of the invariance  of the free-space speed of light in all inertial frames has been  foundational to the physics of the 20$^{th}$ Century, and so to the prevailing physicist's paradigm.  Not only did it provide computational means essential for the standard model of particle physics, but also provided the spacetime ontology,  which  physicists claim to be  one of the greatest of all discoveries, particularly when extended to the current standard model of cosmology, which assumes a curved spacetime account of not only gravity but also of the universe, but necessitating the invention of  dark matter and dark energy. 

 It s usually assumed that the many successes of the resulting Special Theory of Relativity mean that there could be very little reason to doubt the validity of the invariance postulate.  However  the spacetime formalism is just that, a formalism, and one must always be careful in accepting an ontology on the basis of  ill-defined postulates, as in the case of the speed of light, because the postulate never stipulated how the speed of light was to be measured, in particular how clock retardation and length contraction effects were to be corrected.  In contrast to the spacetime formalism Lorentz gave a different neo-Galilean formalism in which space and time were not mixed, but where the special relativity effects were the consequence of absolute motion with respect to a real 3-space.  Recently \cite{CahillMink} the discovery of an exact linear mapping between  the Minkowski-Einstein spacetime class of coordinates and the neo-Galilean class of time and space coordinates was reported.  In the Minkowski-Einstein class the speed of light is invariant by construction, while in the Galilean class the speed is  not invariant.  Hence statements about the speed of light are formalism dependent, and the claim that the successes of SR implies that the speed of light is invariant is bad logic.  So questions about of the speed of light need to be answered by experiments.
 
 There have been many experiments to search for  light speed anisotropy, and they fall generally into two classes - those that successfully detected anisotropy and those that did not.  The reasons for this apparent disparity are now understood,  for  it is important to appreciate that because the speed of light is invariant in SR - as an essential part of that formalism, then SR cannot be used to design or analyse data from light speed anisotropy experiments\footnote{The oft-used Mansouri-Sexl formalism , {\it Gen. Rel. Grav.},  {\bf 8}, 497,1977, is an invalid formalism for analysing anisotropy experiments for it fails to take account of even the refractive index effect in dielectric-mode Michelson interferometers.}. The class of experiments that failed to detect anisotropy, such as those using vacuum Michelson interferometers, say in the form of resonant vacuum cavities \cite{cavities}, suffer a design flaw that was only discovered in 2002 \cite{MMCK,MMC}.  Essentially there is a subtle cancellation effect in the original Michelson interferometer, in that two unrelated effects exactly cancel unless the light passes through a dielectric.  In the original Michelson interferometer experiments the dielectric happened, fortuitously, to be a gas, as in \cite{MM,Miller,Illingworth,Joos,Jaseja,Munera}, and then the sensitivity is reduced by the factor $k^2=n^2-1$, where $n$ is the refractive index of the gas, compared to the sensitivity factor  $k^2=1$ used by Michelson in his calculation of the instrument's calibration constant, using Newtonian physics. For air, with $n=1.00029$,  this factor has value $k^2=0.00058$ which explained why the original Michelson-Morley fringe shifts were much smaller than expected.  The physics that Michelson was unaware of was the reality of the Lorentz-Fitzgerald contraction effect.  Indeed the null results from  the resonant vacuum cavities \cite{cavities} experiments, in comparison with their gas-mode versions, gives explicit proof of the reality of the contraction effect\footnote{As well the null results from the LIGO-like  and related vacuum-mode Michelson interferometers are an even more dramatic confirmation. Note that in contrast the LISA space-based vacuum interferometer does not suffer from the Lorentz contraction effect, and as a consequence would be excessively sensitive.}. A more sensitive and very cheap detector is to use optical fibers as the light carrying medium, as then the cancellation effect is overcome \cite{CahillOF}.  Another technique to detect light speed anisotropy has been to make one-way speed measurements; Torr and Kolen \cite{Torr}, Krisher {\it et a.l} \cite{Krisher}, DeWitte \cite{DeWitte} and Cahill \cite{CahillOFRF}.  Another recently discovered technique is to use the doppler shift data from spacecraft earth-flybys \cite{CahillFlyby}.  Using the spacetime formalism results in an unexplained earth-flyby doppler shift anomaly,  Anderson {\it et al.} \cite{And2008}, simply because the spacetime formalism is one that explicitly specifies  that the speed of the EM waves is invariant, but only wrt a peculiar choice of space and time coordinates.  
 
 Here we combine data from   two high precision NASA/JPL experiments: (i)  the one-way speed of light experiment using optical fibers:   Krisher {\it et al.} \cite{Krisher},  and (ii)  the spacecraft earth-flyby doppler shift data:   Anderson {\it et al.} \cite{And2008}, to  give the solar-system galactic 3-space average speed of   486 km/s in the direction RA=4.29$^h$, Dec=-75$^{\circ}$.  Turbulence effects (gravitational waves) are also evident. Various  data  reveal the 30 km/s orbital speed of the earth and the sun inflow component of 615 km/s near the sun, and  42 km/s at 1AU, and for the first time, experimental evidence of the 3-space inflow of the earth, which is predicted to be 11.2 km/s at the earth's surface.  The optical-fiber and restricted flyby data give, at this stage, only an average of $12.4 \pm 5$ km/s for the earth inflow - averaged over the spacecraft orbits, and so involving averaging wrt distance from earth  and RF propagation angles wrt the inflow\footnote{A spacecraft in an eccentric orbit about the earth would permit, using the high-precision doppler shift technology, a detailed mapping of the 3-space inflow.}.  The optical fiber - flyby data is  in remarkable agreement with the    spatial flow characteristics as determined in other light speed anisotropy experiments, such as Michelson-Morley (1887), Miller (1933), DeWitte (1991), Torr and Kolen (1981),  Cahill (2006), Munera (2007), Cahill and Stokes (2008) and Cahill (2009). The NASA data enables an independent calibration of detectors for use in  light speed anisotropy experiments and related gravitational wave detectors. These are turbulence effects in the flowing 3-space.  These fluctuations are in essence gravitational waves, and which were apparent even in the Michelson-Morley 1887 data \cite{Book,Review,CahillGW2009}.
 
 \section{Flowing 3-Space and Emergent Quantum Gravity}
 We give a brief review of the concept and mathematical formalism of a dynamical flowing 3-space, as this is often confused with the older dualistic space and aether ideas, wherein some particulate aether is located and  moving through an unchanging Euclidean space - here both the space and the aether were viewed as being ontologically real.  The dynamical 3-space is different: here we have only a dynamical 3-space, which at a small scale is a quantum foam system without dimensions and described by fractal  or nested homotopic mappings \cite{Book}.  This quantum foam is not embedded in any space -  the quantum foam is all there is and any metric properties are intrinsic properties solely of that quantum foam. At a macroscopic level the quantum foam is described by a velocity field ${\bf v}({\bf r},t)$, where ${\bf r}$ is merely a $[3]$-coordinate within an embedding space. This space has no ontological existence - it is merely used to (i) record that the quantum foam has, macroscopically, an effective dimension of 3, and (ii) to relate other phenomena also described by fields, at the same point  in the quantum foam.  The dynamics for this 3-space is easily determined by the requirement that observables be independent of the embedding choice, giving, for zero-vorticity  dynamics and for a flat embedding space\footnote{It is easy to re-write (\ref{eqn:3spacedynamics}) for the case of a non-flat embedding space, such as an $S^3$,  by introducing an embedding 3-space-metric $g_{ij}({\bf r})$, in place of the Euclidean metric $\delta_{ij}$. A generalisation of  (\ref{eqn:3spacedynamics}) has also been suggested in \cite{Book} when the vorticity is not zero. This vorticity treatment predicted an additional gyroscope precession effect for the GPB experiment, RT Cahill, {\it Progress in Physics}, {\bf 3}, 13-17, 2007.}
\begin{eqnarray}
\nabla.\left(\frac{\partial {\bf v} }{\partial t}+ ({\bf v}.{\bf \nabla}){\bf v}\right)+
\frac{\alpha}{8}\left((tr D)^2 -tr(D^2)\right)=-4\pi G\rho,  \nonumber 
\end{eqnarray}
\vspace{-6mm}
\begin{eqnarray}
 \nabla\times {\bf v}={\bf 0},  \mbox{\  \  \   }
 D_{ij}=\frac{1}{2}\left(\frac{\partial v_i}{\partial x_j}+
\frac{\partial v_j}{\partial x_i}\right),
\label{eqn:3spacedynamics}\end{eqnarray} 
where $\rho({\bf r},t)$ is the matter and EM energy densities expressed as an effective matter density.  Borehole $g$ measurements and astrophysical blackhole data has shown that $\alpha\approx1/137$ is  the fine structure constant to within observational errors  \cite{Book, Schrod,Review,QC}.  For a quantum system with mass
$m$ the Schr\"{o}dinger equation is  uniquely generalised   \cite{Schrod} with the new terms required to maintain that the motion is intrinsically wrt to the 3-space, and not wrt to the embedding space,  and that the time evolution is unitary
\begin{equation}
i\hbar\frac{\partial \psi({\bf r},t)}{\partial t}  =-\frac{\hbar^2}{2m}\nabla^2\psi({\bf r},t)-i\hbar\left({\bf
v}.\nabla+\frac{1}{2}\nabla.{\bf v}\right) \psi({\bf r},t).
\label{eqn:Schrod}\end{equation}
The space and time coordinates $\{t,x,y,z\}$ in (\ref{eqn:3spacedynamics}) and (\ref{eqn:Schrod}) ensure that  the separation of a deeper and unified process into different classes of phenomena - here a dynamical 3-space (quantum foam) and a quantum matter system, is properly tracked and connected. As well the same coordinates may be used by an observer to also track the different phenomena.  However it is important to realise that these coordinates have no ontological significance - they are not real.  The velocities ${\bf v}$ have no ontological or absolute meaning relative to this coordinate system - that is in fact how one arrives at the form in  (\ref{eqn:Schrod}), and so the ``flow" is always relative to the internal dynamics of the 3-space. 
A quantum wave packet propagation analysis of (\ref{eqn:Schrod}) gives  the acceleration induced by wave refraction to be \cite{Schrod}
\begin{equation}
{\bf g}=\frac{\partial{\bf v}}{\partial t}+({\bf v}.\nabla){\bf v}+
(\nabla\times{\bf v})\times{\bf v}_R,
\label{eqn:acceln}\end{equation}
 \vspace{-4mm}
\begin{equation}
{\bf v}_R({\bf r}_0(t),t) ={\bf v}_0(t) - {\bf v}({\bf r}_0(t),t),
\end{equation}
where ${\bf v}_R$ is the velocity of the wave packet relative to the 3-space, and where ${\bf v}_O$ and ${\bf r}_O$ are the velocity and position relative to the observer, and the last term in (\ref{eqn:acceln}) generates the Lense-Thirring effect as a vorticity driven effect.  Together (\ref{eqn:Schrod}) and (\ref{eqn:acceln}) amount to the derivation of gravity as a quantum effect,  explaining  both the  equivalence principle ($\bf g$ in (\ref{eqn:acceln}) is independent of $m$) and the Lense-Thirring effect. Overall we see, on ignoring vorticity effects, that
\begin{equation}
\nabla.{\bf g}=-4\pi G\rho-\frac{\alpha}{8}\left((tr D)^2 -tr(D^2)\right),
\label{eqn:NGplus}\end{equation}
which is Newtonian gravity but with the extra dynamical term whose strength is given by $\alpha$. This new dynamical effect explains the spiral galaxy flat rotation curves  (and so doing away with the need for ``dark matter"), the bore hole $g$ anomalies,  the black hole ``mass spectrum". Eqn.(\ref{eqn:3spacedynamics}), even when $\rho=0$, has an expanding universe Hubble solution that fits the recent supernovae data in a parameter-free manner without requiring ``dark matter" nor ``dark energy", and without the accelerating expansion artifact \cite{QC,Paradigm}. However (\ref{eqn:NGplus}) cannot be entirely expressed in terms  of ${\bf g}$ because the fundamental dynamical variable is $\bf v$. The role of  (\ref{eqn:NGplus}) is to reveal that if we analyse gravitational phenomena we will usually find that the matter density $\rho$ is insufficient to account for the observed ${\bf g}$. Until recently this failure of Newtonian gravity has been explained away as being caused by some unknown and undetected ``dark matter" density.  Eqn.(\ref{eqn:NGplus}) shows that to the contrary it is a dynamical property of 3-space itself. Here we determine various properties of this dynamical 3-space from the NASA optical-fiber and spacecraft flyby doppler anomaly data.

Significantly the quantum matter 3-space-induced  `gravitational' acceleration in (\ref{eqn:acceln}) also follows from minimising the elapsed proper time wrt the wave-packet trajectory ${\bf r}_o(t)$, see \cite{Book},
\begin{equation}
\tau=\int dt \sqrt{1-\frac{{\bf v}^2_R({\bf r}_0(t),t)}{c^2}}
\label{eqn:propertime}\end{equation}
and then taking the limit $v_R/c \rightarrow 0$. This shows that (i) the matter `gravitational' geodesic is a quantum wave refraction effect, with the trajectory determined by a Fermat least proper-time  principle, and (ii) that quantum systems undergo a local time dilation effect - which is used later herein in connection with the  Pound-Rebka experiment. A full derivation of 
(\ref{eqn:propertime}) requires the generalised Dirac equation.

\section{3-Space Flow Characteristics and the Velocity Superposition Approximation}  
This paper reports the most detailed analysis so far of data from various experiments that have directly detected the 3-space velocity field  ${\bf v}({\bf r},t)$.  The dynamics in  (\ref{eqn:3spacedynamics}) is necessarily time-dependent  and having various contributing effects, and in order of magnitude: (i) galactic flows associated with the motion of the solar system within the Milky Way, as well as  flows caused by the supermassive black hole at the galactic center and flows associated with the local galactic cluster, (ii) flows caused by the orbital motion of the earth and of the inflow caused by the Sun, and (iii) the inflow associated with the earth. An even smaller flow associated with the moon is not included in the analysis. It is necessary to have some expectations of the characteristics of the flow expected for an earth based observer. First consider an isolated spherical mass density $\rho(r)$,  with total mass $M$, then
(\ref{eqn:3spacedynamics}) has stationary flow solution, for $r>R$, i.e outside of the mass,
\begin{equation}
{\bf v}({\bf r})=-\hat{{\bf r}}\sqrt{\frac{2GM(1+\frac{\alpha}{2}+..)}{r}}
\label{eqn:sphericalflow}\end{equation}
which gives the matter acceleration from   (\ref{eqn:acceln}) to be
\begin{equation}
{\bf g}({\bf r})=-\hat{{\bf r}}\frac{GM(1+\frac{\alpha}{2}+..)}{r^2}
\label{eqn:sphericalaccel}\end{equation}
corresponding  to a gravitational potential, via $\bf g= -\nabla\Phi$,
\begin{equation}
\Phi({\bf r})=-\frac{GM(1+\frac{\alpha}{2}+..)}{r}
\label{eqn:sphericalpotential}\end{equation}
This special case is Newton's law of gravity, but with some 0.4$\%$  of the effective mass being caused by the $\alpha$- dynamics term. The inflow (\ref{eqn:sphericalflow}) would be applicable to an isolated and stationary sun or  earth.  At the surface of the sun this predicts an inflow speed of $615$ km/s, and $42$ km/s at the earth distance of 1AU. For the earth itself the inflow speed at the earth's surface is predicted to be $11.2$ km/s.   When both occur and when both are moving wrt the asymptotic 3-space, then  numerical solutions of  (\ref{eqn:3spacedynamics})  are required.  However an approximation that appears to work is to assume that the net flow in this case may be approximated by a vector superposition \cite{CahillSuper2009}
 \begin{equation}
 {\bf v}={\bf v}_{galactic}+{\bf v}_{sun}-{\bf v}_{orbital}+{\bf v}_{earth}+...
 \label{eqn:superposition}\end{equation}
 which are, in order, translational motion of the sun, inflow into the sun, orbital motion of the earth (the orbital motion produces an apparent flow in the opposite direction - hence the -ve sign; see Fig.\ref{fig:orbit}), inflow into the earth, etc. The first three have been previously determined from experimental data, and here we more accurately and using new data determine all of these components.  However this superposition cannot completely valid as  (\ref{eqn:3spacedynamics}) is non-linear. So  the superposition may be at best approximately valid as a time average only.  The experimental data has always shown that the detected flow is time dependent, as one would expect, as with multi-centred mass distributions no stationary flows are known.  This time-dependence is a turbulence effect - it is in fact easily observed and is seen in the Michelson-Morley 1887 data \cite{Review}. This turbulence is caused by the presence of any significant mass, such as the galaxy, sun, earth. The  NASA/JPL data discussed herein again  displays  very apparent turbulence.  These wave effects are essentially {\it gravitational waves}, though they have characteristics different from those predicted from GR, and have a different interpretation.  Nevertheless for a given flow ${\bf v}({\bf r},t)$, one can determine the corresponding induced  spacetime metric $g_{\mu\nu}$ which generates the same matter geodesics as from (\ref{eqn:NGplus}), with the proviso that this metric is not determined by the Hilbert-Einstein equations of GR.  Significantly vacuum-mode Michelson interferometers cannot detect this phenomenon, which is why LIGO and related detectors have not seen these very large wave effects.    
 
 \section{Gas-Mode Michelson Interferometer}
 The Michelson interferometer is a brilliant instrument for measuring ${\bf v}({\bf r},t)$, but only when operated in dielectric mode. This is because two different and independent effects exactly cancel in vacuum mode; see \cite{Book, Review,MMCK}. Taking account of the geometrical path differences, the Fitzgerald-Lorentz arm-length contraction and  the Fresnel drag effect leads to the travel time difference between the two arms, and which is detected by interference effects\footnote{The dielectric of course does not cause the observed effect, it is merely a necessary part of the instrument design physics, just as mercury in a thermometer does not {\it cause} temperature.}, is given by 
\begin{figure}
\setlength{\unitlength}{0.75mm}
\hspace{30mm}\begin{picture}(0,30)
\thicklines

\definecolor{hellgrau}{gray}{.8}
\definecolor{dunkelblau}{rgb}{0, 0, .9}
\definecolor{roetlich}{rgb}{1, .7, .7}
\definecolor{dunkelmagenta}{rgb}{.9, 0, .0}
\definecolor{green}{rgb}{0, 1,0.4}
\definecolor{black}{rgb}{0, 0, 0}
\definecolor{dunkelmagenta}{rgb}{.9, 0, .0}

 \color{dunkelmagenta}
  \put(10,5){\vector(0,1){5}}
\put(11,25){\vector(0,-1){5}}
\put(11,30){\line(0,-1){38}}
\put(11,-2){\vector(0,-1){5}}
\put(-10,0){\line(1,0){50}}
\put(-5,0){\vector(1,0){5}}
\put(40,-1){\line(-1,0){29.2}}
\put(15,0){\vector(1,0){5}}
\put(30,-1){\vector(-1,0){5}}
\put(10,0){\line(0,1){30}}


 \color{black}
 \put(9,-8){\line(1,0){5}}
\put(9,-9){\line(1,0){5}}
\put(14,-9){\line(0,1){1}}
\put(9,-9){\line(0,1){1}}

\put(6.5,30){\line(1,0){8}}
\put(40,-4.5){\line(0,1){8}}
 \put(40,-4.5){\line(0,1){8}}

 \put(8.0,-2){\line(1,1){5}}
\put(9.0,-2.9){\line(1,1){5}}
 \put(35,-5){ $B$}
\put(25,-5){ $L$}
\put(12,26){ $C$}
 \put(4,-5){ $A$}
 \put(15,-9){ $D$}

\put(5,12){ $L$}

 \color{dunkelmagenta}
\put(50,0){\line(1,0){50}}
 \put(55,0){\vector(1,0){5}}
\put(73,0){\vector(1,0){5}}
\put(85,0){\vector(1,0){5}}
 
\put(100,-1){\vector(-1,0){5}}

\put(68.5,-1.5){\line(1,1){4}}
\put(69.3,-2.0){\line(1,1){4}}

\put(70,0){\line(1,4){7.5}}
\put(70,0){\vector(1,4){3.5}}
\put(77.5,30){\line(1,-4){9.63}}
\put(77.5,30){\vector(1,-4){5}}

\put(83.3,-1.5){\line(1,1){4}}
\put(84.0,-2.0){\line(1,1){4}}
\put(100,-1){\line(-1,0){14.9}}

 \color{black}
\put(73.5,30){\line(1,0){8}} 
\put(100,-4.5){\line(0,1){8}}
\put(82.3,-2.5){\line(1,1){4}}
 \put(82,-2){\line(1,1){5}}
\put(83,-2.9){\line(1,1){5}}

 \put(85,-8){\line(1,0){5}}
\put(85,-9){\line(1,0){5}}
\put(90,-9){\line(0,1){1}}
\put(85,-9){\line(0,1){1}}

 \put(67,-5){ $A_1$}
\put(80,-5){ $A_2$}
\put(90,-9){ $D$}
\put(95,-5){ $B$}
\put(79,26){ $C$}
\put(85,16){ $v_P$}
\put(-8,8){(a)}
\put(55,8){(b)}
\put(60,15){gas}
\put(25,15){gas}

\color{dunkelblau}
\put(85,10){\vector(2,1){15}}  
\put(85,10){\line(1,0){12}}  
\put(93,11){$\theta-\psi$}  

\end{picture}

\vspace{10mm}
\caption{\small{Schematic diagrams of the gas-mode Michelson Interferometer, with beamsplitter/mirror at $A$ and
mirrors at $B$ and $C$ mounted on arms  from $A$, with the arms of equal length $L$ when at rest.  $D$ is the detector screen. In (a) the interferometer is at rest in space. In (b) the 3-space  is moving through the gas and the interferometer with speed $v_P$ in the plane of the interferometer and direction  $\delta=\theta-\psi$ relative to $AB$ arm. Interference fringes are observed at  $D$ when mirrors $C$ and $D$ are not exactly perpendicular.  As the interferometer is rotated in the plane  shifts of the fringes are seen in the case of absolute motion, but only if the apparatus operates in a gas.  By measuring fringe shifts the speed $v_P$ may be determined. In general the $v_P$ direction has angle $\theta$ wrt the local meridian, and the arm $AB$ has angle $\psi$ relative to the local meridian, so that $\delta=\theta-\psi$ is angle between ${\bf v}_P$ and one-arm. The difference in travel times $\Delta t$ is given in  (\ref{eqn:e5}), but with temperature changes and non-orthogonal mirrors by (\ref{eqn:e6}). In vacuum the fringes do not shift during rotation.}  \label{fig:Minterferometer}}
\end{figure}
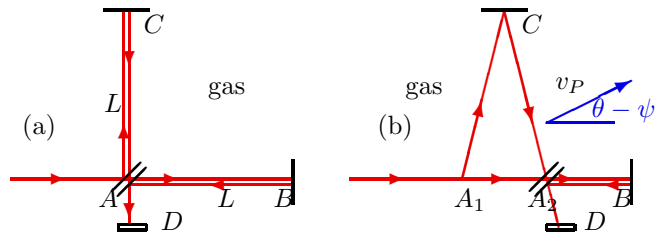

 \begin{equation}
\Delta t=k^2\frac{Lv_P^2}{c^3}\cos
\bigl(2 (\theta-\psi)\bigr) ,
\label{eqn:e5}\end{equation}
where $\psi$ specifies the direction of   ${\bf v}({\bf r},t)$  projected onto the plane of the interferometer, giving projected value $v_P$, relative to the local meridian, and where $k^2= (n^2-2)(n^2-1)/n$. Neglect of the relativistic Fitz\-gerald-Lorentz contraction effect
gives   $k^2\,{\approx}\, n^3\,{\approx }\,1$
 for gases, which is essentially the Newtonian theory that Mich\-elson used.
 
 However the above analysis does not correspond to how the interferometer is actually operated. That analysis does not actually predict fringe shifts, for the  field of view would be uniformly illuminated, and the observed effect would be a changing level of luminosity rather than fringe shifts. As Michelson and Miller knew, the mirrors must be made slightly non-orthogonal with the degree of non-orthogonality determining how many fringe shifts were visible in the field of view. Miller exper\-i\-mented with this effect to determine a comfortable number of fringes: not too few and not too many.  Hicks \cite{Hicks} deve\-lop\-ed a theory for this effect -- however it is not necessary to be aware of the details of this analysis in using the interferometer: the non-orthogonality   reduces the symmetry of the device, and instead of having period of 180$^\circ$ the symmetry now has a period of 360$^\circ$, so that to (\ref{eqn:e5})  we must add the extra term $a\cos(\theta-\beta)$ in
\vspace*{-4pt}
 \begin{equation}
\Delta t=k^2\frac{L(1+e\theta)v_P^2}{c^3}\cos\bigl(2(\theta-\psi)\bigr)+a(1+e\theta)\cos(\theta-\beta)+f
\label{eqn:e6}\end{equation}   
The term $1+e\theta$ models the temperature effects, namely that as the arms are uniformly rotated, one rotation taking several minutes, there will be a temperature induced change in the length of the arms. If the temperature effects are linear in time, as they would be for short time intervals, then they are linear in $\theta$. In the  Hick's term the parameter $a$ is proportional to the length of the arms, and so also has the temperature factor. The term $f$ simply models any offset  effect.
Michelson and Morley and Miller took these two effects  into account  when analysing his data. The Hick's effect is particularly apparent in the Miller  and Michelson-Morley data.

\begin{figure}
\vspace{0mm}
\hspace{30mm}\includegraphics[scale=0.26]{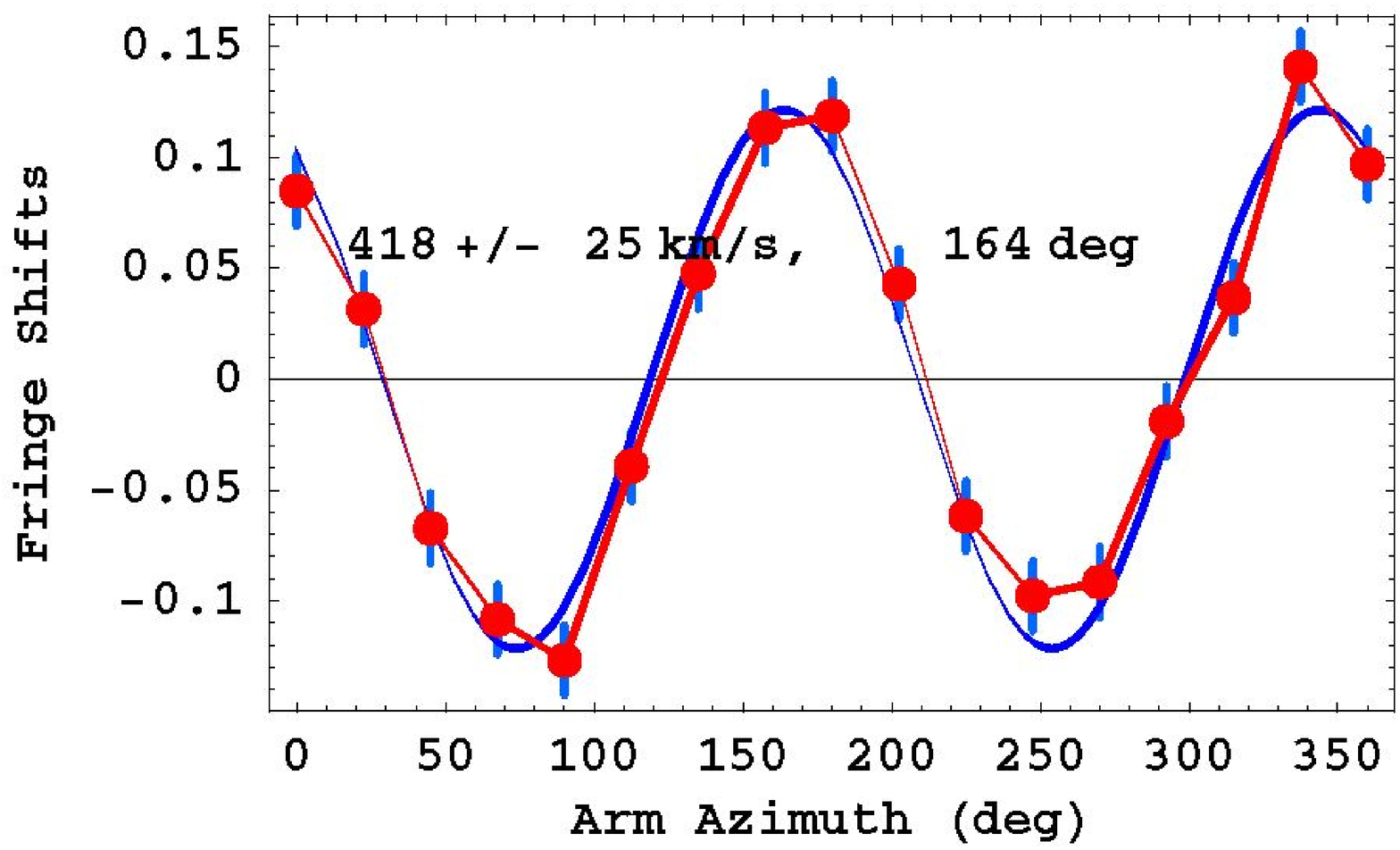}

\hspace{30mm}\includegraphics[scale=0.26]{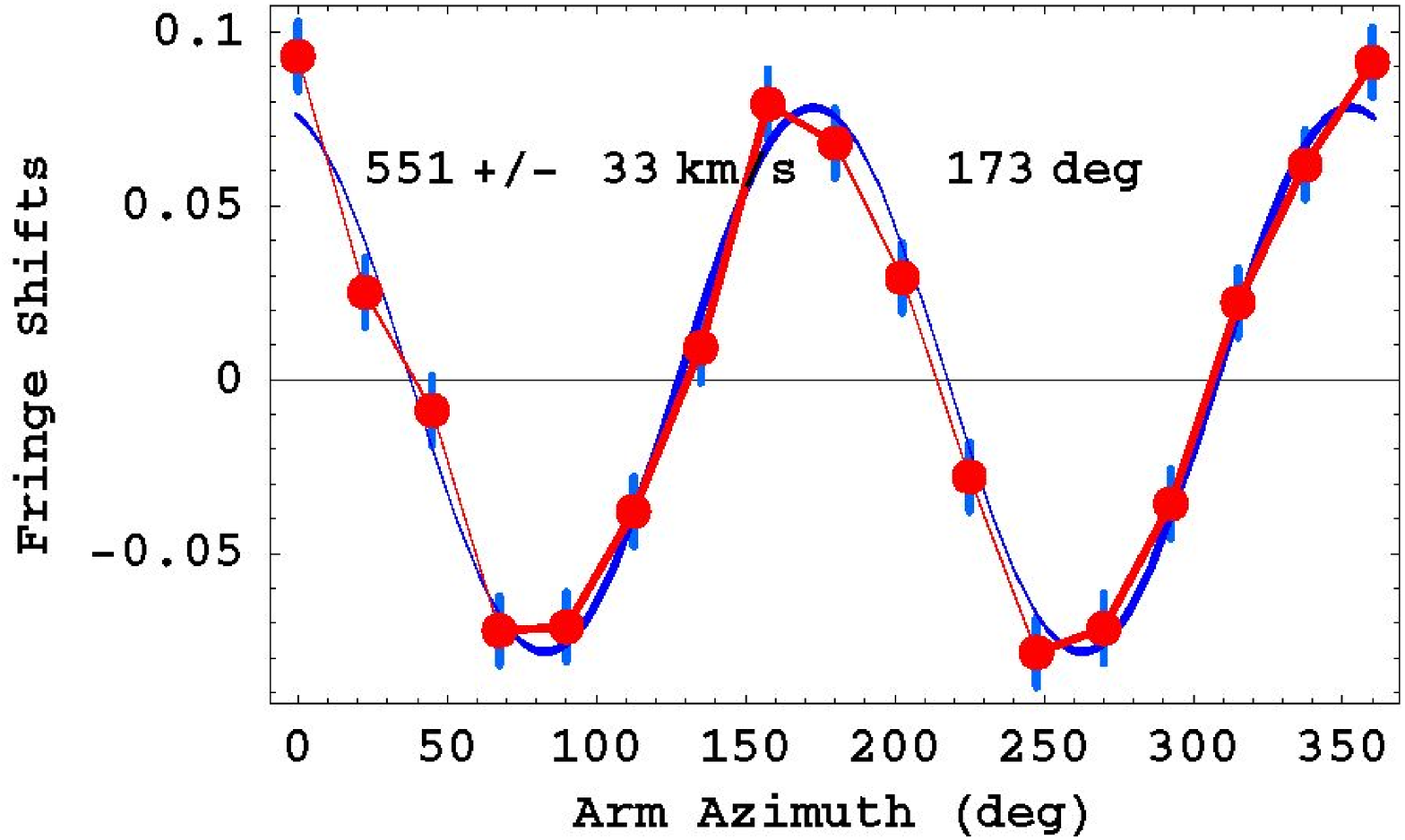}

\caption{\small {(a) A typical  Miller  averaged-data from September 16, 1925, $4^h 40^\prime$  Local Sidereal Time (LST) - an average of data from 20 turns of the gas-mode Michelson interferometer. Plot and data after fitting using  (\ref{eqn:e6}), and then subtracting both the temperature drift and Hicks effects from both, leaving the expected sinusoidal form.  The error bars are determined as the rms error in this fitting procedure, and show how exceptionally small were the errors, and which agree with Miller's claim for the errors. (b) Best result  from the Michelson-Morley 1887 data - an average of 6 turns, at  $7^h$  LST on July  11, 1887.  Again the rms error is remarkably small.  In both cases the indicated speed is  $v_P$ - the 3-space speed projected onto the plane of the interferometer. The angle is  the azimuth of the  3-space speed projection at the particular LST.  The speed fluctuations from day to day significantly exceed these errors, and reveal the existence of 3-space flow turbulence - i.e gravitational waves.}}
\label{fig:MillerMMPlots}\end{figure}

The interferometers are operated with the arms horizon\-tal. Then in (\ref{eqn:e6}) $\theta$ is the azimuth of one arm
relative to the local meridian, while $\psi$ is the azimuth of the absolute motion
velocity projected onto the plane of the interferometer, with projected component $v_P$.
Here the Fitzgerald-Lorentz con\-traction is a real dynamical effect of absolute motion,
unlike the Einstein spacetime view that it is merely a spacetime perspective artifact, and
whose magnitude depends on the choice of observer. The instrument is operated by rotating
at a rate of one rotation over several minutes, and observing the shift in the fringe
pattern through a telescope during the rotation.  Then fringe shifts from six (Michelson
and Morley) or twenty (Miller) successive rotations are averaged to improve the signal to noise ratio, and the average sidereal time noted. Some examples are shown in Fig.\ref{fig:MillerMMPlots}, and illustrate the incredibly clear signal.  The ongoing claim that the Michelson-Morely experiment was a null experiment is disproved.  And as well, as discussed in \cite{Book,Review,CahillGW2009}, they detected gravitational waves, {\it viz} 3-space turbulence in 1887. The new data analysed herein is from one-way optical fiber and doppler shift spacecraft experiments. The agreement between these and the gas-mode interferometer techniques demonstrate that the Fitzgerald-Lorentz contraction effect is a real dynamical effect.  The null results from the vacuum-mode interferometers  \cite{cavities} and  LIGO  follow simply from having $n=1$ giving $k^2=0$ in (\ref{eqn:e5}).

\begin{figure}[t]
\vspace{0mm}
\hspace{40mm}\includegraphics[scale=1.1]{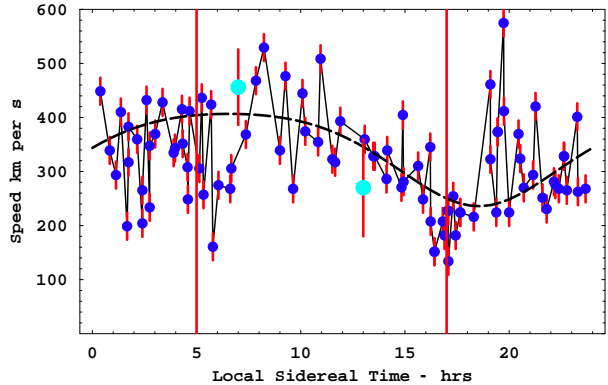}
\caption{\small {Speeds $v_P$, of the 3-space velocity ${\bf v}$  projected onto the horizontal plane of the  Miller gas-mode Michelson interferometer located atop 
Mt.Wilson,  plotted against local sidereal time in hours,  for a composite day, with data  collected over a number of days in September 1925, \cite{Miller}.    The data shows considerable fluctuations, from hour to hour, and also day to day, as this is a composite day.  The dashed curve shows the non-fluctuating  best-fit variation over one day, as the earth rotates, causing the projection onto  the plane of the interferometer of the  velocity of the average direction of the space flow to change.  The maximum projected speed of the curve is  $417$ km/s (using the STP air refractive index of $n=1.00029$ in (\ref{eqn:e5}) (atop Mt. Wilson the better value of $n=1.00026$ is suggested by the NASA data), and the  min/max occur at approximately 5hrs and 17hrs local sidereal time (Right Ascension).  Note from Fig.\ref{fig:CelestialPlot} and Table \ref{table:flyby}   that the Cassini flyby in August gives a RA$= 5.15^h$, close to the RA apparent in the above plot. The error bars are determined by the method discussed in Fig.\ref{fig:MillerMMPlots}. The green data points, with error bars, at $5^h$ and $13^h$ are from the Michelson-Morley  1887 data, from averaging (excluding only the July 8 data for 7$^h$ because it has poor S/N), and with same rms error analysis. The fiducial time lines at $5^h$ and $17^h$  are the same as those shown in Figs.\ref{fig:phases} and \ref{fig:CelestialPlot}. The speed fluctuations are seen to be much larger than the statistically determined errors,  confirming the presence of  turbulence in the 3-space flow, i.e gravitational waves, as first seen in the Michelson-Morley experiment. }}
\label{fig:Miller}\end{figure}

\section[Sun 3-Space Inflow from  Miller Interferometer]{Sun 3-Space Inflow from  Miller Interferometer  Data\label{section:suninflow}}
 
\begin{table*}
\hspace{5mm}
\hspace{0mm}\begin{tabular}{|l|l|c|l|l|l|l|} 
\hline 
{\bf Epoch 1925/26} &\mbox{\ \ }$v_M$ & $\overline{k}$ &$ v=v_M/k_{air}$  & $\overline v=v_M/\overline{k}$ 
&$v=\sqrt{3}\overline{v}$&\mbox{\ \  }
$v_{sun}$\\
\hline\hline  
 February 8  &9.3 km/s & 0.048 & 385.9 km/s  & 193.8 km/s & 335.7 km/s &  51.7 km/s \\ \hline
 April 1  & 10.1 &0.051 &419.1  & 198.0 &342.9  &56.0 \\ \hline 
 August 1  & 11.2 &0.053 &464.7  & 211.3 &366.0  &58.8 \\ \hline
 September 15  & 9.6 &0.046 &398.3  & 208.7 &361.5  & 48.8\\ \hline
\hline
\end{tabular}

\caption{\small  The $\overline{k}$ anomaly:  $\overline{k} \gg k_{air}=0.0241$, as  the
3-space inflow effect. Here $v_M$ and $\overline{k}$ come from fitting the interferometer data using Newtonian physics (with $v_{orbital}=30$ km/s used to determine 
 $\overline{k}$), while $\,v\,$ and $\,\overline{v}\,$  are computed speeds using the indicated scaling. The  average of the sun inflow speeds, at 1AU, is $v_{sun}=54\pm6$ km/s, compared to the  predicted inflow speed of $42$ km/s from (\ref{eqn:sphericalflow}).
From column 4 we obtain the average galactic flow of $v=417\pm50$ km/s, compared with the NASA-data determined flow of $486$ km/s. } 
\label{table:sunflow}
\end{table*}

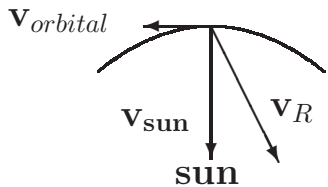
\begin{figure}[ht]
\vspace{10mm}

\setlength{\unitlength}{0.6mm}

\hspace{60mm}\begin{picture}(20,-20)
\thicklines
\put(25,10){\vector(-1,0){15}}
\put(25,10){\vector(0,-1){29.5}}
\put(25,10){\vector(1,-2){15}}
\qbezier(0,0)(25,20)(50,0)
\put(5,-12){\Large $\bf v_{sun}$}
\put(17,-25){\Large $\bf sun$}
\put(-20,10){\Large ${\bf v}_{orbital}$}
\put(38,-10){\Large ${\bf v}_{R}$}
\end{picture}
\vspace{15mm} 
\caption{\small  Orbit of earth about the sun defining the  plane of the ecliptic with tangential orbital
velocity ${\bf v}_{orbit}$ and the sun inflow velocity  ${\bf v}_{sun}$. Then ${\bf v}_{R}={\bf v}_{sun}-{\bf v}_{orbit}$ is the velocity of the 3-space relative to the earth, but not showing the
${\bf v}_{galactic}$ contribution. }
\label{fig:orbit}
\end{figure}

Miller was led to the conclusion that for reasons unknown  the existing theory of
the Michelson interferometer did not reveal true values of $v_P$, and for this reason  he introduced the parameter $k$,  with $\overline{k}$ herein indicating his numerical values. Miller had reasoned  that he could determine both ${\bf v}_{galactic}$ and $\overline{k}$ by observing the
interferometer- determined $v_P$ and $\psi$ over a year because the known orbital speed  of the earth about the sun of $30$ km/s would modulate both of these observables, giving what he termed an aberration effect as shown in Fig.\ref{fig:CelestialPlot}, and by a scaling argument he could determine the absolute
velocity of the solar system.   In this manner he finally determined that $|{\bf v}_{galactic}|=208$ km/s in the direction $(\alpha=4^h 54^m, \delta=-70^0 33^\prime)$.  However now that the theory of the Michelson interferometer has been revealed an anomaly becomes apparent.  Table 3 shows $v=v_M/k_{air}$, the speed determined using (\ref{eqn:e5}), for each of the four epochs.   However Table
3 also shows that $\overline{k}$ and the speeds $\overline{v}=v_M/\overline{k}$ determined by the
scaling argument are considerably different.  We denote by $v_M$ the notional speeds determined from (\ref{eqn:e5}) using the Michelson Newtonian-physics value of $k=1$. The $v_M$ values arise after taking account of  the projection effect. That  $\overline{k}$ is considerably larger than the value of
$k_{air}$ indicates that  another velocity component has been overlooked.   Miller of course only knew of
the tangential  orbital speed of the earth, whereas the new physics predicts that as-well there is a
3-space radial inflow ${\bf v}_{sun}=42$ km/s at 1AU. We can approximately re-analyse  Miller's
data to extract a first approximation to the speed of this inflow component.   Clearly it is
$v_R=\sqrt{v_{sun}^2+v^2_{orbital}}$ that sets the scale, see Fig.\ref{fig:orbit} and not
$v_{orbital}$, and because
$\overline{k}=v_M/v_{orbit}$ and $k_{air}=v_M/v_R$ are the scaling relations, then 
\begin{eqnarray}\label{eqn:QG9}
v_{sun}&=&v_{orbital}\sqrt{\displaystyle{ \frac{v_R^2}{v_{orbital}^2}-1 }},  \nonumber \\
      &=&v_{orbital}\sqrt{\displaystyle{ \frac{\overline{k}^2}{k_{air}^2}-1 }}.  
\end{eqnarray} 
Using the  $\overline{k}$  values in Table \ref{table:sunflow} and the value\footnote{We have not modified this value to take account of the altitude effect  or temperatures atop  Mt.Wilson. This weather information was not recorded by Miller. The temperature and pressure effect  is that $n=1.0+0.00029\frac{P}{P_0}\frac{T_0}{T}$, where $T$ is the temperature in $^0$K and $P$ is the pressure in atmospheres.  $T_0=273$K  and $P_0=$1atm. The NASA data implies that atop Mt. Wilson the air refractive index was probably close to $n=1.00026$. } of $k_{air}$  we obtain the $v_{sun}$ speeds  shown in Table \ref{table:sunflow}, which give an averaged speed of $54\pm6$ km/s, compared to the predicted inflow speed of $42$ km/s.   Of course this simple re-scaling of the Miller results is not completely valid because  the direction of ${\bf v}_R$ is of course different to that of ${\bf v}_{orbital}$, nevertheless the sun inflow speed of $v_{sun}=54\pm5$ km/s at 1AU from this analysis is reasonably close to the predicted value of $42$ km/s.

\section{Generalised Maxwell Equations and the Sun 3-Space Inflow Light Bending}\label{sect:maxwell}

\begin{figure}
\setlength{\unitlength}{2.0mm}
\vspace{0mm}
\hspace{40mm}\begin{picture}(10,30)
\thicklines

\put(5,25){\line(1,0){12}}   
\put(10,25){\vector(1,0){1}}
\put(23,24){\line(4,-1){12}}
\put(23,24){\vector(4,-1){7}}
\put(23,25){\line(1,0){12}}
\put(27,23.0){ $\delta$}
\qbezier(17,25)(20,24.8)(23,24)  

\put(20,20){\line(1,0){10}}  
\put(20,20){\line(-1,0){10}}  
\put(20,20){\line(0,1){10}}  
\put(20,20){\line(0,-1){10}}  
\put(20,20){\line(1,1){7}}  
\put(20,20){\line(1,-1){7}}  
\put(20,20){\line(-1,1){7}}  
\put(20,20){\line(-1,-1){7}}  

\put(13,20){\vector(1,0){1}}
\put(27,20){\vector(-1,0){1}}
\put(20,27){\vector(0,-1){1}}
\put(20,13){\vector(0,1){1}}
\put(25,25){\vector(-1,-1){1}}
\put(25,15){\vector(-1,1){1}}
\put(15,15){\vector(1,1){1}}
\put(15.5,24.5){\vector(1,-1){1}}

\put(20,20){\circle{9}}  

\end{picture}
\vspace{-23mm}\caption{\small{ Shows bending of light through angle $\delta$ by the inhomogeneous spatial inflow, according to the minimisation of the travel time in (\ref{eqn:lighttime}). This effect permits the inflow speed at the surface of the sun to be determined to be $615$ km/s. The inflow speed into the sun at the distance of the earth from the sun has been extracted from the Miller data, giving $54\pm6$ km/s. } }\label{fig:light}
\end{figure}
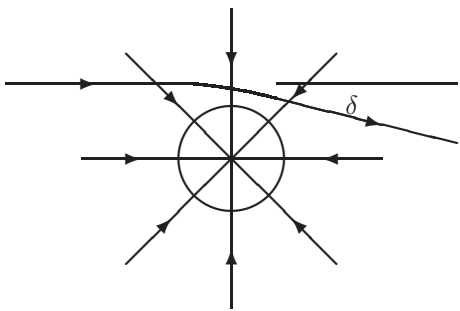
One of the putative key tests of the GR formalism was the gravitational bending of light by the sun during the 1915 solar eclipse.  However this effect also immediately follows from the new 3-space dynamics once we also generalise the Maxwell equations so that the electric and magnetic  fields are excitations of the dynamical space. The dynamics of the electric and magnetic fields  must then have the form, in empty space,  
$$
\displaystyle{ \nabla \times {\bf E}=-\mu\left(\frac{\partial {\bf H}}{\partial t}+{\bf v.\nabla H}\right)},
\displaystyle{ \nabla \times {\bf H}=\epsilon\left(\frac{\partial {\bf E}}{\partial t}+{\bf v.\nabla E}\right)} 
$$
\begin{equation}
\displaystyle{\nabla.{\bf H}={\bf 0}} ,   \mbox{\ \ \ \  }
\displaystyle{\nabla.{\bf E}={\bf 0}}
\label{eqn:MaxE18a}\end{equation}
which was first suggested by Hertz in 1890  \cite{Hertz}, but with $\bf v$ being a constant vector field. 
Suppose we have a uniform flow of space with velocity ${\bf v}$ wrt the embedding space or wrt an observer's frame of reference. Then we can find plane wave solutions for (\ref{eqn:MaxE18a}):
\begin{equation}
{\bf E}({\bf r},t)={\bf E}_0e^{i({\bf k}.{\bf r}-\omega t)} \mbox{\ \ \ \  } {\bf H}({\bf r},t)={\bf H}_0e^{i({\bf k}.{\bf r}-\omega t)}
\label{eqn:pw}\end{equation}
with
\begin{equation}
\omega({\bf k},{\bf v})=c|\vec{{\bf k}}| +{\bf v}.{\bf k} \mbox{ \ \ \  where \ \ \  } c=1/\sqrt{\mu\epsilon}
\label{eqn:omega}\end{equation}
Then the EM group velocity is
\begin{equation}
{\bf v}_{EM}=\vec{\nabla}_k\omega({\bf k},{\bf v})=c\hat{\bf k}+{\bf v}
\label{eqn:groupv}\end{equation}
So the velocity of EM radiation ${\bf v}_{EM}$ has magnitude  $c$ only with respect to the space, and in general not with respect to the observer if the observer is moving through space. These experiments show that the speed of light is in general anisotropic, as predicted by (\ref{eqn:groupv}). The time-dependent and inhomogeneous  velocity field causes the refraction of EM radiation. This can be computed by using the Fermat least-time approximation. Then the EM ray paths  ${\bf r}(t)$ are determined by minimising  the elapsed travel time:
\begin{equation}
\tau=\int_{s_i}^{s_f}\frac{ds\displaystyle{|\frac{d{\bf r}}{ds}|}}{|c\hat{{\bf v}}_R(s)+{\bf v}(\bf{r}(s),t(s)|},
\label{eqn:lighttime}\end{equation}
\begin{equation}
{\bf v}_R=\left(  \frac{d{\bf r}}{dt}-{\bf v}(\bf{r}(t),t)\right),
\label{eqn:vR}\end{equation}
by varying both ${\bf r}(s)$ and $t(s)$, finally giving ${\bf r}(t)$. Here $s$ is a path parameter, and ${\bf v}_R$ is the 3-space   vector tangential to the path. For light bending by the sun inflow (\ref{eqn:sphericalflow}) the angle of deflection  is
\begin{equation}
\delta=2\frac{v^2}{c^2}=\frac{4GM(1+\frac{\alpha}{2}+..)}{c^2d}+...
\label{eqn:E19}\end{equation}
where $v$ is the inflow speed at distance $d$  and $d$ is the impact parameter. This agrees with the GR result except for the $\alpha$ correction.  Hence the  observed deflection of $8.4\times10^{-6}$ radians is actually a measure of the inflow speed at the sun's surface, and that gives $v=615$ km/s, in agreement with (\ref{eqn:sphericalflow}).   These generalised Maxwell equations also predict gravitational lensing produced by the large inflows associated with the new `black holes' in galaxies.

 \begin{figure*}
\vspace{-25mm}
\hspace{34mm}\includegraphics[scale=1]{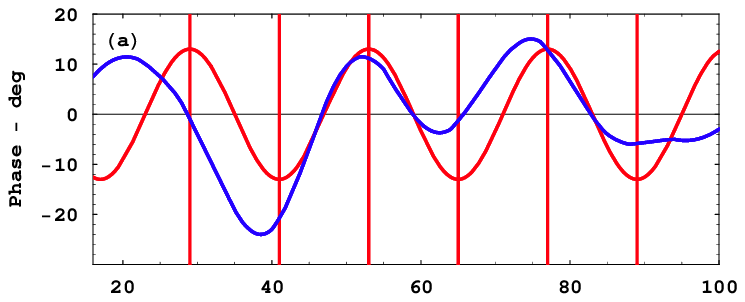}

\hspace{34mm}\includegraphics[scale=1]{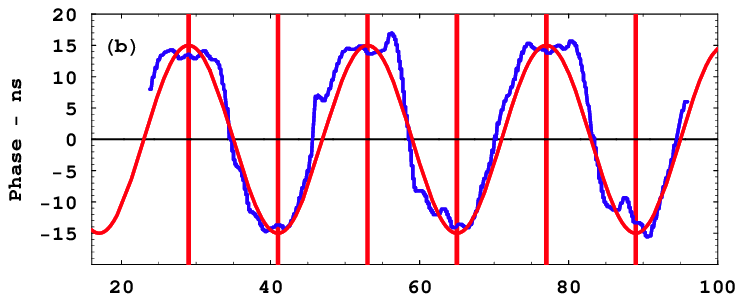}

\hspace{35mm}\includegraphics[scale=0.98]{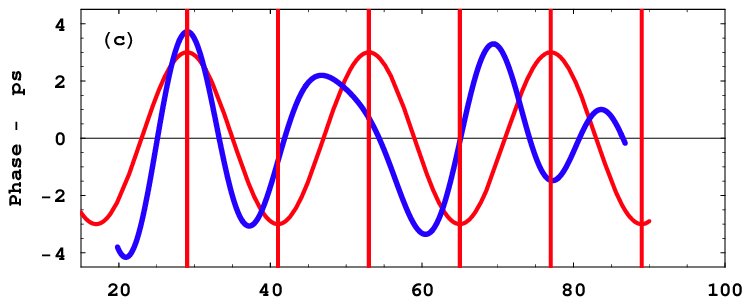}

\hspace{31.5mm}\includegraphics[scale=1.01]{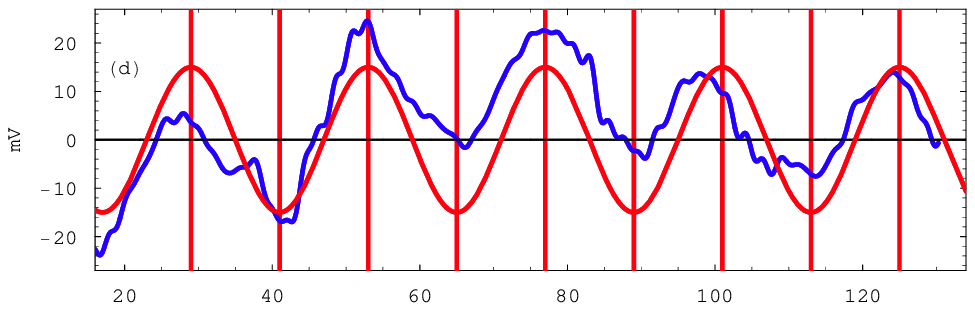}

\hspace{34mm}\includegraphics[scale=0.892]{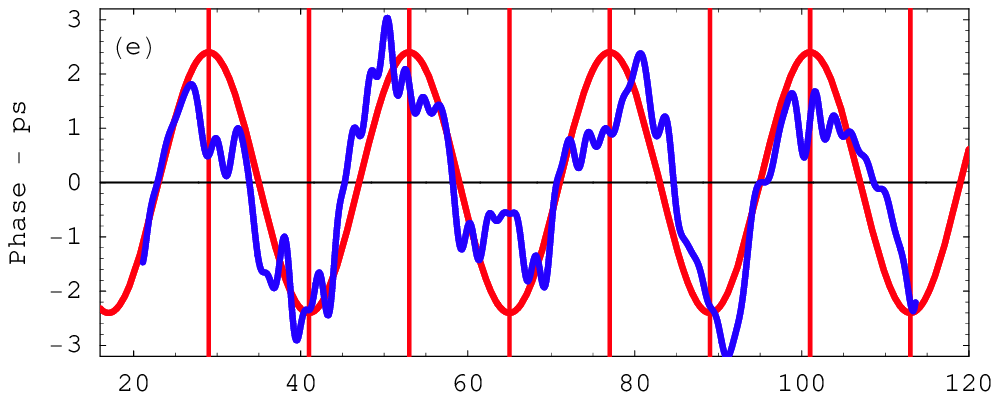}

\hspace{22.5mm}\includegraphics[scale=1.16]{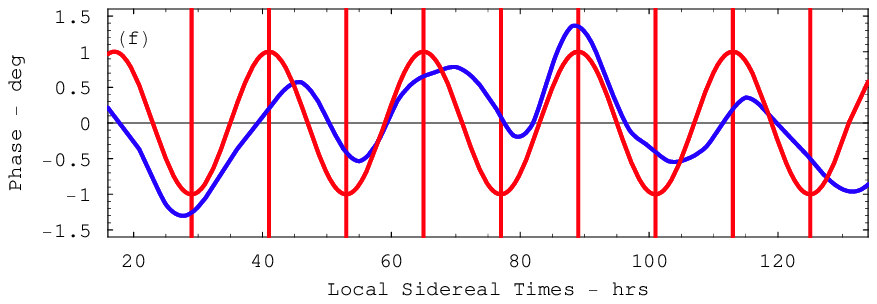}

\vspace{-4mm}\caption{\small{Data from five different EM speed anisotropy experiments showing earth rotation wrt local preferred frame, as shown by sidereal time phasing, together with wave effects.  In all cases a zero bias was removed and low-pass filtering was applied. {\bf (a)}: Krisher \cite{Krisher} optical fiber phase difference data $\phi_1-\phi_2$, in degrees.    {\bf (b)}: DeWitte \cite{DeWitte} RF coaxial cable phase data, in ns. The DeWitte cable ran NS.  {\bf (c):} Cahill \cite{CahillOFRF} hybrid optical-fiber/RF coaxial-cable data, in ps, from August 2006. Cable ran NS.  {\bf (d):} Cahill  \cite{CahillOF, CahillStokesOF} optical-fiber Michelson interferometer, in photodiode mV, from September 18, 2007.   {\bf (e):} Cahill \cite{CahillFresnel} RF coaxial-cable data, in ps, from May 2009. Cable ran NS. {\bf (f)}: Krisher \cite{Krisher} optical fiber phase sum data $\phi_1+\phi_2$, in degrees. In each case the (red) sinusoidal curves shows the phase expected for a RA of $5^h$, but with arbitrary magnitudes. The vertical lines are at local sidereal times of 5$^h$ and 17$^h$, on successive days, corresponding to the RAs shown in red in Fig.\ref{fig:CelestialPlot}. The Krisher data gives a local sidereal time of $4.96^h$, corresponding to  a RA of $6.09^h$ for November - caused by the 42$^\circ$ inclination of the optical fiber to the local meridian.  This RA was used in combination with the spacecraft  earth-flyby doppler shift data.  Note the amplitude and phase fluctuations in  all the data - these are gravitational wave effects. }}
\label{fig:phases}\end{figure*}

 \section{Torr and Kolen RF One-Way Coaxial Cable Experiment}
 
 \begin{figure}
\vspace{0mm}
\hspace{40mm}\includegraphics[scale=1.0]{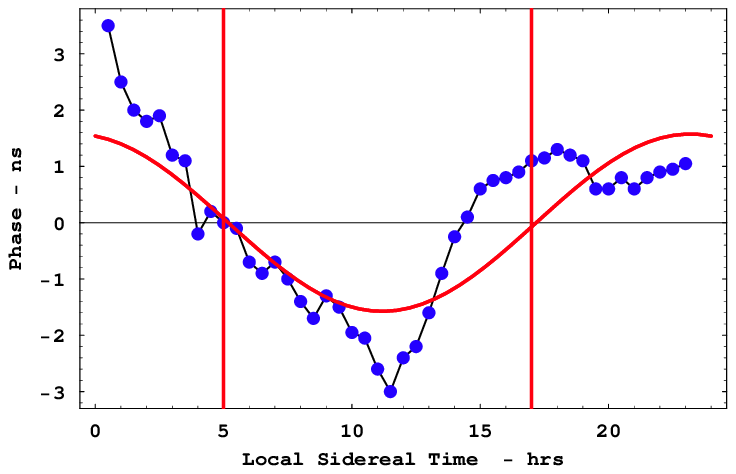}
\vspace{-2mm}
\caption{\small{Data from the 1981 Torr-Kolen experiment at Logan, Utah \cite{Torr}. 
The data shows variations in travel times
(ns),  for local sidereal times,  of an RF signal travelling through $500$ m of coaxial 
cable  orientated in an EW direction. Actual days are not indicated but the experiment 
was done during February-June 1981.   Results are for  a typical day.   For the 1st of
February the  local time of $12\!\!:\!\!00$  corresponds to $13\!\!:\!\!00$   sidereal
 time.  The predictions are for   February, 
for a cosmic speed of $480$ km/s in the direction  ($\alpha=5.0^h, \delta=-70^0$), and including orbital and in-flow velocities but without theoretical turbulence.  The vertical lines are at local sidereal times of 5$^h$ and 17$^h$, corresponding to the RAs shown in red in Figs. \ref{fig:phases} and \ref{fig:CelestialPlot}. }}
\label{fig:TorrKolen}\end{figure}

A one-way coaxial cable\index{RF coaxial cable!Torr-Kolen} experiment  was
performed at the Utah University in 1981 by Torr and  Kolen \cite{Torr}. This involved two rubidium vapor clocks placed approximately $500$ m apart with a 5 MHz sinewave RF signal propagating between the clocks via a nitrogen filled coaxial cable buried in the ground and maintained at a constant pressure of $\sim$2 psi. Torr and Kolen  observed  variations in the one-way travel time, as shown in Fig.\ref{fig:TorrKolen} by the  data points.   The theoretical  predictions for the  Torr-Kolen experiment for a cosmic speed of $480$ km/s in the direction  ($\alpha=5^h, \delta=-70^\circ$), and including orbital and in-flow velocities, are  shown in Fig.\ref{fig:TorrKolen}.   The maximum/minimum effects occurred, typically,  at the predicted times.    Torr and Kolen reported fluctuations in both the magnitude, from 1 - 3 ns, and time of the maximum variations in travel time, just as  observed in all  later experiments - namely wave effects.

\section{Krisher {\it et al.}  One-Way Optical-Fiber Experiment}

The Krisher {\it et al. } one-way  experiment \cite{Krisher}  used two hydrogen- maser oscillators  with light sent in each direction through optical fiber of length approximately 29 km.   The optical fiber was part of the NASA DSN Deep Space Communications Complex in the Mojave desert at Goldstone, California.  Each maser provided a stable 100-MHz output frequency. This signal was split, with one signal being fed directly into one channel of a Hewlett-Packard Network Analyzer. The other signal was used to modulate a laser carrier signal propagated along a $29$ km long ultrastable fiber optics link that is buried five feet underground. This signal was fed into the second channel of the other Network Analyzer at the distant site. Each analyzer is used to measure the relative phases of the masers, $\phi_1$ and $\phi_2$. The data collection began on November 12 1988 at 20:00:00 (UTC), with phase measurements made every ten seconds until November 17 1988 at 17:30:40 (UTC).  Figs.\ref{fig:phases}(a) and (f) shows plots of the phase difference $\phi_1 - \phi_2$ and phase sum $\phi_1 + \phi_2$, in degrees,  after removing a bias and a linear trend, as well as being filtered using a Fast Fourier Transform.  The data is plotted against local sidereal time. In analysing the phase data the propagation path was taken to be along a straight line between the two masers, whose longitude and latitude are given by $(243^\circ 12^\prime21^{\prime\prime}.65,$ $ 35^\circ 25^\prime33^{\prime\prime}.37)$ and  $(243^\circ 06^\prime40^{\prime\prime}.37,$ $  35^\circ 14^\prime51^{\prime\prime}.82)$. Fig.\ref{fig:phases} 
     shows as well  the corresponding phase differences from other experiments.    Krisher only compared the phase variations with that  of the Cosmic Microwave Background (CMB), and noted that the phase relative to the local sidereal time differed from   CMB direction by 6 hrs, but failed to notice that it agreed with the direction discovered by Miller in 1925/26 and published in 1933 \cite{Miller}.  The phases from the various experiments show that, despite very different longitudes of the experiments and  different days in the year, they are in phase when plotted against local sidereal times. This demonstrates that the phase cycles are caused by the rotation of the earth relative to the stars - that we are observing a galactic phenomenon, being that the 3-space flow direction is 
reasonably steady  wrt the galaxy\footnote{The same effect is observed in Ring Lasers \cite{RingLaser} - which detect a  sidereal period of rotation of the earth, and not the solar period. Ring Lasers cannot detect the 3-space direction, only a relative  rotation angle.}. Nevertheless we note that all the phase data show fluctuations in both the local sidereal time for maxima/minima and also fluctuations in magnitude. These wave effects first  appeared in experimental data of Michelson and Morley in  1887.

From the November  Krisher data in Figs.\ref{fig:phases}(a) and (f) the Right Ascension of the 3-space flow direction  was obtained from the local sidereal times of the maxima and minima, giving a RA  of  $6.09^h$, after correcting  the  apparent RA of $4.96^h$ for  the 42$^\circ$ inclination of the optical fiber to the local meridian.  This RA was used in combination with the spacecraft  earth-flyby doppler shift data, and is shown in Fig.\ref{fig:CelestialPlot}. 

The magnitudes of the Krisher phases are not used in determining the RA for November, and so are not directly used in this report.  Nevertheless these magnitudes provide a check on the physics of how the speed of light in  optical fibers is affected by the 3-space flow. The phase differences $\phi_1-\phi_2$ in Fig.\ref{fig:phases}a, which correspond to a 1st order in $v/c$ experiment in which the Fresnel drag effect must be taken into account,  are shown to be consistent in \cite{CahillFresnel}   with the determined speed   for November, noting that   the use of phase comparators does not allow the determination of  multiple $360^\circ$ contributions to the phase differences.  The analysis of the  Krisher phase sum $\phi_1+\phi_2$ in Fig.\ref{fig:phases}f, which correspond to a 2nd order in $v/c$ experiment, requires  the Lorentz contraction of the optical fibers. as well as the  Fresnel drag effect, to be taken into account.  The physics of optical fibers in relation to this and other 3-space physics is discussed more fully in Cahill \cite{CahillFresnel}.

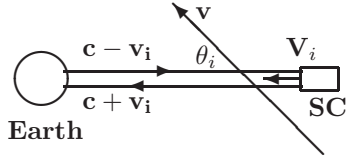
\begin{figure}
\vspace{20mm}
\hspace{3mm}
\setlength{\unitlength}{1.0mm}
\hspace{40mm}\begin{picture}(0,0)
\thicklines
\put(9.5,5){\circle{7.0}}
\put(5,-3){\bf Earth}
\put(12.5,6){\line(1,0){31.7}}
\put(12.5,4){\line(1,0){31.7}}
\put(24,6.0){\vector(1,0){3}}
\put(24,4.0){\vector(-1,0){3}}

\put(44,3.5){\line(1,0){5}}
\put(44,6.5){\line(1,0){5}}
\put(49,3.5){\line(0,1){3}}
\put(44,3.5){\line(0,1){3}}
\put(45,0.0){\bf SC}

\put(47,-5){\vector(-1,1){20}} 

\put(44,5){\vector(-1,0){5}} 

\put(42,8){${\bf   V}_i$} 

\put(15,8){$\bf    c-v_i$}

\put(15,1){$\bf    c+v_i$}
\put(30,13){$\bf    v$}
\put(30,7){$\theta_i$}

\end{picture}
\vspace{5mm}
\caption{ \small {Asymptotic flyby configuration in earth frame-of-reference, with spacecraft (SC) approaching Earth with velocity ${\bf V}_i$. The departing asymptotic velocity will have a different direction but the same speed, as no force other than conventional Newtonian gravity is assumed to be acting upon the SC. The dynamical 3-space velocity is ${\bf v}({\bf r},t)$, though taken to be time independent during the doppler shift measurement, which causes the outward EM beam  to have speed $c-v_i(r)$, and inward speed  $c+v_i(r)$, where $v_i(r)=v(r)\cos(\theta_i)$, with $\theta_i$ the angle between $\bf v$ and $\bf V$.  A similar description applies to the departing SC, labeled $i \rightarrow f$. }}
 \label{fig:doppler}
\end{figure}
 
 \begin{figure}[t]
\vspace{27mm}
\hspace{0mm}
\setlength{\unitlength}{1.0mm}
\hspace{65mm}\begin{picture}(0,0)
\thicklines
\put(6.5,5){\circle{20.0}}
\put(2.6,4.5){\bf Earth}
\put(13.5,5){\line(1,5){2.1}}
\put(16,3){\bf TS}
\put(14.0,6){\circle{2.0}}

\put(16,13){\bf SC}
\put(11,14){$ \theta_i$}
\put(13,19){ $\bf v(D)$}
\qbezier(-5,6)(3,14)(14,16)
\put(15,16.1){\line(5,1){15}}
\put(14.5,16.5){\vector(-1,+1){4}}
\put(15,16){\circle*{2.0}}

\put(-5,6){\line(-1,-1){10}}
\put(15,9){{  D}}
\put(26,18.5){\vector(-4,-1){2}}
\put(25,14){${\bf V}_i$}

\put(-10.9,0){\vector(-1,-1){2}}
\put(-12,-6){${\bf V}_f$}

\end{picture}
\vspace{5mm}
\caption{\small{ Spacecraft (SC) earth flyby trajectory, with initial and final asymptotic velocity ${\bf V}$, differing only by direction. The doppler shift is determined from Fig.\ref{fig:doppler} and (\ref{eqn:newspeedi}).  The 3-space flow velocity at the location of the SC is  ${\bf v}$. The line joining Tracking Station (TS) to  $SC$ is the path of the RF signals, with length $D$.  As SC approaches earth ${\bf v}(D)$ changes direction and magnitude, and hence magnitude of projection $v_i(D)$ also changes,  due to earth component of 3-space flow and also because of RF direction to/from Tracking Station. The SC trajectory averaged magnitude of this earth in-flow is determined from the flyby data and compared with theoretical prediction.} }
 \label{fig:flyby}
\end{figure}
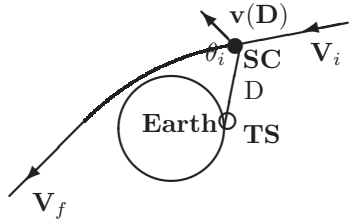

\section{3-Space Flow from  Earth-Flyby Doppler Shifts}

The motion of  spacecraft relative to the earth are measured by observing the direction   and doppler shift of the   transponded RF transmissions.  This gives another  technique to  determine the speed and direction of the dynamical 3-space as  manifested by the light speed anisotropy \cite{CahillFlyby}. The repeated detection of the anisotropy of the speed of light has been, until recently, ignored in analysing the doppler shift data, causing the long-standing anomalies in the analysis   \cite{And2008}.  The use of the Minkowski-Einstein choice of time and space coordinates does not permit the analysis of these doppler anomalies, as they mandate that the speed of the EM waves be invariant. 

Because we shall be extracting the earth inflow effect we need to take account of  a spatially varying, but not time-varying,  3-space velocity.
In the earth frame of reference, see Fig.\ref{fig:doppler}, and using clock times from earth-based clocks, let the transmitted signal from earth  have frequency $f$.  The time for one RF maximum to travel distance $D$  to SC from earth  is, see Fig.\ref{fig:flyby}, 
\begin{equation}
t_1=\int_0^D \frac{dr}{c-v_i(r)}
\end{equation}
The next RF maximum leaves  time $T=1/f$ later and arrives at SC at time, taking account of SC motion,
\begin{equation}
t_2=T+\int_0^{D-VT} \frac{dr}{c-v_i(r)}
\end{equation}
The period at the SC of the arriving RF is then
\begin{equation}
T^\prime=t_2-t_1=T+\int_D^{D-VT} \frac{dr}{c-v_i(r)}\approx \frac{c-v_i(D)-V}{c-v_i(D)}T
\label{eqn:period}\end{equation}
Essentially this RF is reflected\footnote{In practice a more complex protocol is used.}
 by the SC. Then the 1st RF maximum takes time to reach the earth
 \begin{equation}
t_1^{\prime}=-\int^0_{D-VT} \frac{dr}{c+v_i(r)}
\end{equation}
and the 2nd RF maximum takes time
\begin{equation}
t_2^{\prime}=T^\prime-\int^0_{D-VT-VT^\prime} \frac{dr}{c+v_i(r)}.
\end{equation}
Then the period of the returning RF at the earth is 
\begin{eqnarray}
T^{\prime\prime}&=&t_2^{\prime}-t_1^{\prime}\nonumber \\
&=&T^\prime+\int_{D-VT}^{D-VT-VT^\prime} \frac{dr}{c+v_i(r)}\nonumber \\
&\approx& \frac{c+v_i(D-VT)-V}{c+v_i(D-VT)}T^\prime \nonumber \\
&\approx& \frac{c+v_i(D)-V}{c+v_i(D)}T^\prime
\end{eqnarray}
Then overall we obtain the return frequency to be\footnote{This corrects the corresponding expression in \cite{CahillFlyby}, but without affecting the final results.} 
\begin{equation}
f^{\prime\prime}=\frac{1}{T^{\prime\prime}}=\frac{c+v_i(D)}{c+v_i(D)-V}.\frac{c-v_i(D)}{c-v_i(D)-V}f
\label{eqn:frequency}
\end{equation}
Ignoring the projected 3-space velocity $ v_i(D)$, that is, assuming that the speed of light is invariant as per the usual literal interpretation of the Einstein 1905 light speed postulate,   we obtain instead
\begin{equation}
f^{\prime\prime}=\frac{c^2}{(c-V)^2}f.
\label{eqn:oldfrequency}
\end{equation}
The use of  (\ref{eqn:oldfrequency}) instead of  (\ref{eqn:frequency})  is the origin of the putative anomalies.    Expanding  (\ref{eqn:oldfrequency})
we obtain
\begin{equation}
\frac{\Delta f}{f}=\frac{ f^{\prime\prime}-f}{f}=\frac{2V}{c}
\label{eqn:olddopplera}
\end{equation}
However expanding (\ref{eqn:frequency}) we obtain, for the same doppler shift,
\begin{equation}
\frac{\Delta f}{f}=\frac{ f^{\prime\prime}-f}{f}=\left(1+\frac{v(D)^2}{c^2}\right)\frac{2V}{c}+ .... 
\label{eqn:newspeed}\end{equation}
It is the  prefactor  to $2V/c$ missing from (\ref{eqn:olddopplera}) that explains the spacecraft doppler anomalies, and also permits yet another  determination of the 3-space velocity  ${\bf v}(D)$, {\it viz} at the location of the SC. The published data does not give the doppler shifts as a function of SC location, so the best we can do at present is to use a SC 
trajectory-averaged $v(D)$, namely $\overline{v}_i$ and $\overline{v}_f$, for the incoming and outgoing trajectories, as further discussed below.

\begin{figure}
\vspace{0mm}
\hspace{40mm}\includegraphics[scale=1.1]{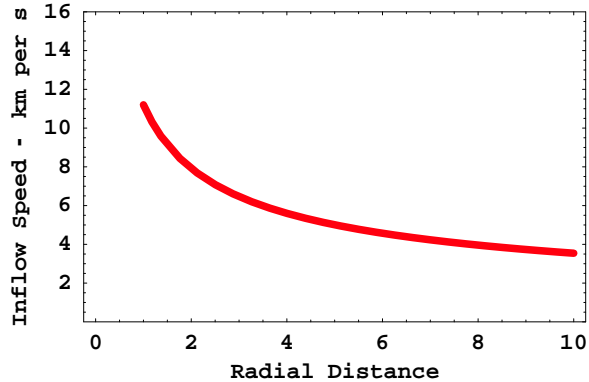}
\caption{Earth 3-space inflow speed vs distance from earth in earth radii, as given in (\ref{eqn:sphericalflow}), plotted only for $R>1.0$. Combining the NASA/JPL optical fiber RA determination and the flyby doppler shift data has permitted the determination of the angle- and distance-averaged inflow speed, to be $12.4\pm5$km/s.}
\label{fig:EarthSpeed}\end{figure}

\begin{figure*}
\vspace{-25mm}
\hspace{12mm}\includegraphics[scale=1.8]{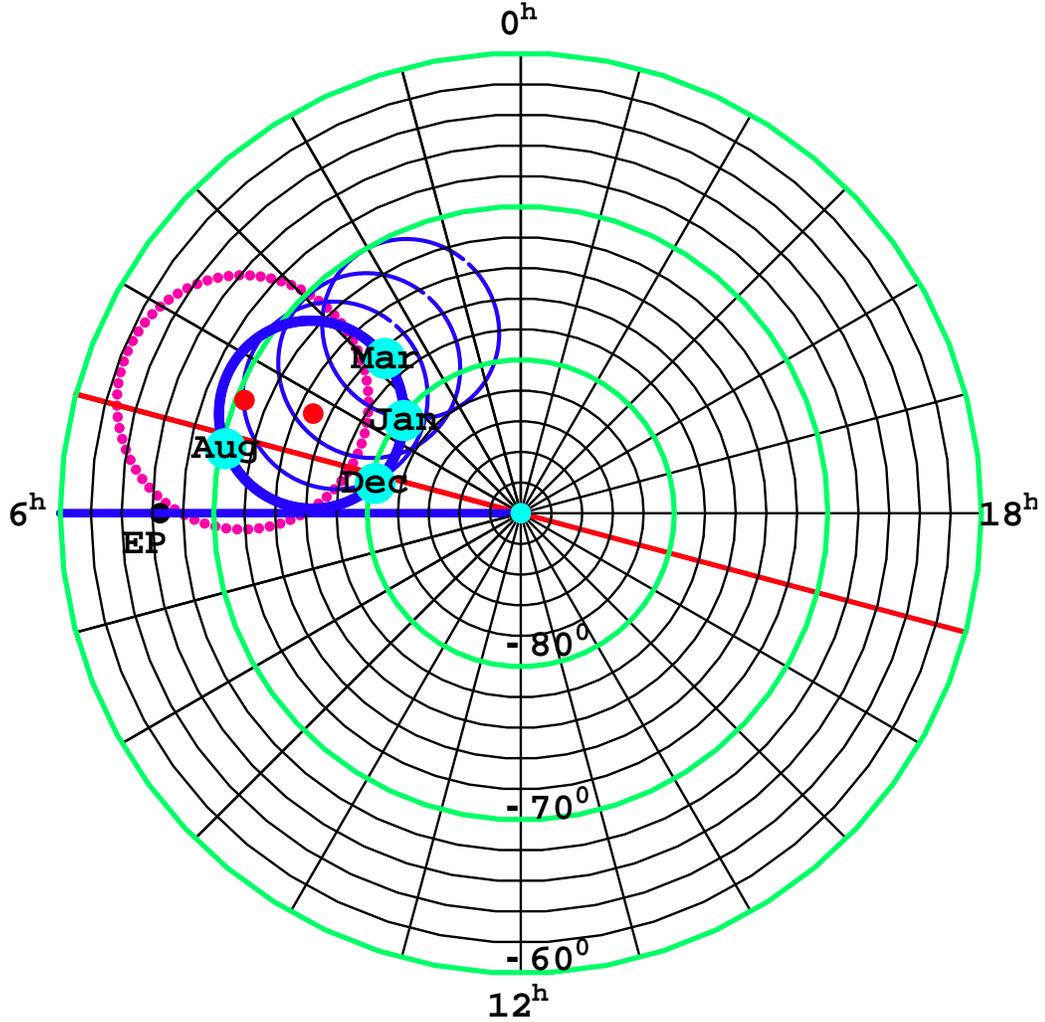}
\vspace{0mm}
\caption{  \small South celestial sphere with RA and Dec shown. The red dotted circle shows the  Miller aberration path discovered in 1925/26, from \cite{Miller}.  The red point at $\alpha=4.52$hrs, $\delta=-70.5^\circ$ shows the galactic flow direction determined by Miller, after removing earth-orbit aberration effect.   The dark blue circle shows the aberration path from best-fitting the earth-flyby doppler shift data and using the optical-fiber RA data point for November from Krisher \cite{Krisher}, see Fig.\ref{fig:KFlybyPlot}. This corresponds to a best fit averaged earth inflow speed of $12.4\pm5$km/s. The blue aberration paths show the best-fit if (a) upper circle:   earth inflow speed =  0 km/s, (b)  = 4.0 km/s, (c) = 8.0 km/s and (d) = 12.4 km/s (thick blue circle).  The actual 3-space flow directions  are shown by light-blue background to   labels  for the flybys in Aug, Dec, Jan and Mar, and given in Table \ref{table:flyby}.   The red point at $\alpha=4.29$hrs, $\delta=-75.0^\circ$ shows the optical-fiber/earth-flyby  determined galactic flow direction, also after removal of earth-orbit aberration effect, and is only $5^\circ$ from the above mentioned Miller direction. The miss-fit angle $\Delta \theta$ between the best-fit RA and Dec for each flyby is given in  Table \ref{table:flyby}, and are only a few degrees on average, indicating the high precision of the fit. This plot shows the remarkable concordance between the NASA/JPL determined 3-space flow characteristics  and those determined by Miller in 1925/26. It must be emphasised that the optical-fiber/flyby aberration  plot and galactic 3-space flow direction is obtained completely independently of the Miller data. The blue line at $56.09^h$ is the orientation corrected  Krisher RA, and has an uncertainty of $\pm 1^h$, caused by wave/turbulence effects. The fiducial RA of 5$^h$ and 17$^h$, shown in red, are the fiducial local sidereal times shown in Figs. \ref{fig:Miller}, \ref{fig:phases} and \ref{fig:TorrKolen}. The point EP  is the pole of the ecliptic. The speed and declination differences between the Miller and NASA data  arise from Miller being unaware of the sun 3-space inflow effect - correcting for this and using an  air refractive index of $n=1.00026$ atop Mt. Wilson increases the Miller data determined speed and moves the declination slightly southward, giving  an even better agreement with the NASA data.   Here we have merely reproduced the Miller aberration plot from \cite{Miller}.}
  \label{fig:CelestialPlot}
\end{figure*}

\begin{figure}[t]
\vspace{0mm}
\hspace{30mm}\includegraphics[scale=1.1]{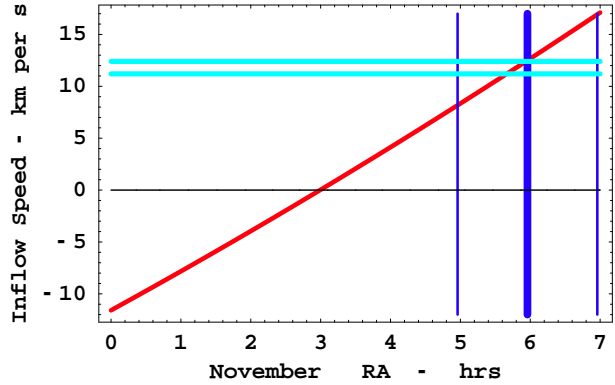}
\caption{\small {The weighted angle- and distance-averaged earth 3-space inflow speed $v_{earth}$, see Fig.\ref{fig:EarthSpeed}, as determined from NASA data, upper green plot. Uses the averaged Right Ascension from the Krisher {\it et al.} data for November, $\alpha=4.96^h$,  but corrected to $\alpha=6.09^h$ for orientation effect of the optical fiber, shown by the thick blue line, with uncertainty range from wave effects  shown by two thin blue lines,  compared with the predicted RA from fitting the flyby data, as shown in Fig.\ref{fig:CelestialPlot}}. The red plot shows that prediction for various averaged inflow speeds, with +ve speeds being an inflow, while -ve speeds are an outflow.  The earth flyby aberration fits for $v_{earth}=0, +4.0$,  $+8.0$ and $+12.4$ km/s are shown in Fig.\ref{fig:CelestialPlot}. Theory gives that  the inflow speed is $+11.2$km/s at the earth's surface - shown by lower green plot. So the detected averaged inflow speed seems to be in good agreement with an expected averaged value.  This is the first detection of the earth's spatial inflow, and the acceleration of this flow is responsible  for the earth's gravity. Note that the flyby data clearly mandates an inflow (+ve values in this figure and not an out-flow - having -ve values).}
\label{fig:KFlybyPlot}\end{figure}

From the observed doppler shift data acquired during a flyby, and then  best fitting the trajectory,  the asymptotic hyperbolic speeds   $V_{i\infty}$ and  $V_{f\infty}$ are inferred from (\ref{eqn:olddopplera}),  but incorrectly so, as in  \cite{And2008}. These inferred asymptotic speeds may be related to an inferred asymptotic doppler shift
\begin{equation}
\frac{\Delta f_{i\infty}}{f}=\frac{ f^{\prime}_\infty-f}{f}=\frac{2V_{i\infty}}{c}+..
\label{eqn:olddoppler}
\end{equation}
which from (\ref{eqn:newspeed}) gives 
\begin{equation}
V_{i\infty}\equiv \frac{\Delta f_{i\infty}}{f}.\frac{c}{2}=\left(1+\frac{\overline{v}_i^2}{c^2}\right)V+ .... 
\label{eqn:newspeedi}\end{equation}
where $V$ is the actual asymptotic speed. 
Similarly after the flyby we obtain
\begin{equation}
V_{f\infty}\equiv \frac{\Delta f_{f\infty}}{f}.\frac{c}{2}=\left(1+\frac{\overline{v}_f^2}{c^2}\right)V+ .... 
\label{eqn:newspeedf}\end{equation}
and we see that the ``asymptotic" speeds   $V_{i\infty}$ and $V_{f\infty}$  must differ, as indeed reported in
 \cite{And2008}.  We then obtain the expression for the so-called flyby anomaly
\begin{equation}
\Delta V_\infty = V_{f\infty}- V_{i\infty} =\frac{\overline{v}_f^2-\overline{v}_i^2}{c^2}V\label{eqn:anomaly}
\label{eqn:speed}\end{equation}
where here $V \approx V_\infty$ to sufficient accuracy, where $ V_\infty$  is the average of  $V_{i\infty}$ and $V_{f\infty}$, 
The existing data on $\bf v$ permits {\it ab initio} predictions for  $\Delta V_\infty$. As well a separate least-squares-fit to the individual flybys permits the determination of the average speed and direction of the 3-space velocity, relative to the earth, during each flyby. These results are all remarkably consistent with  the data from the various laboratory experiments that studied  $\bf v$.  We now indicate how $\overline{v}_i$ and  $\overline{v}_f$ were parametrised during the best-fit to the flyby data. In (\ref{eqn:superposition})  ${\bf v}_{galactic}+{\bf v}_{sun}-{\bf v}_{orbital}$ was taken as constant during each individual flyby,  with  ${\bf v}_{sun}$ inward towards the sun, with value $42$ km/s, and ${\bf v}_{orbital}$ as tangential to earth orbit with value $30$ km/s - consequentially the directions of these two vectors changed with day of each flyby.  The earth inflow ${\bf v}_{earth}$ in (\ref{eqn:superposition})  was taken as radial and of an  unknown  fixed trajectory-averaged value.  So the averaged direction but not the averaged speed  varied from flyby to flyby, with the incoming  and final direction being approximated by the ($\alpha_i, \delta_i$) and ($\alpha_f, \delta_f$) asymptotic directions shown in Table \ref{table:flyby}. The predicted theoretical   variation of  $ v_{earth}(R)$ is shown in Fig.\ref{fig:EarthSpeed}.  To best constrain the fits to the data the flyby data was used in conjunction with the RA from the Krisher optical fiber data.  This results in the aberration plot in Fig.\ref{fig:CelestialPlot}, the various flyby data in Table.\ref{table:flyby}, and the earth in-flow speed determination in Fig.\ref{fig:KFlybyPlot}. The results are in remarkable agreement with the results from Miller, showing the extraordinary skill displayed by Miller in carrying out his massive interferometer experiment and data analysis in 1925/26. The only effect missing from the Miller analysis is the spatial in-flow effect  into the sun, which affected his data analysis, but which has been partially corrected for in Sect.\ref{section:suninflow}.  Miller obtained a galactic flow direction of $\alpha=4.52$ hrs, $\delta=-70.5^\circ$, compared to that obtained herein from the NASA data of  $\alpha=4.29$ hrs, $\delta=-75.0^\circ$, which differ by only $\approx 5^\circ$.

\begin{table*}
\hspace{-6mm}
{
\begin{tabular}{lcccccc}
\hline\hline
Parameter & GLL-I  & GLL-II & NEAR & Cassini & Rosetta & M'GER \\
\hline
Date  &  Dec 8, 1990 &  Dec 8, 1992 &  Jan 23, 1998 & Aug 18, 1999  & Mar 4, 2005  & Aug 2, 2005   \\

$V_\infty$ km/s &  8.949 & 8.877  &  6.851 & 16.010 & 3.863  &  4.056  \\
$\alpha_i$ deg  & 266.76  & 219.35  &  261.17 & 334.31  & 346.12  &  292.61 \\
$\delta_i$ deg  & -12.52  & -34.26  &-20.76   &  -12.92 & -2.81  &  31.44 \\
$\alpha_f$ deg  &219.97   & 174.35  &  183.49 & 352.54  &   246.51&   227.17 \\
$\delta_f$ deg  &-34.15& -4.87  &  -71.96 & -4.99 & -34.29  & -31.92  \\

\hline
$\alpha_v$ hrs &5.23   & 5.23&  3.44 & 5.18 &  2.75 &  4.89\\
$\delta_v$ deg  &-80.3& -80.3 &  -80.3 & -70.3 & -76.6 & -69.5  \\
$v$ km/s  &490.6& 490.6  &  497.3 & 478.3 & 499.2 &479.2 \\
\hline
(O) $\Delta V_\infty$ mm/s  & 3.92$\pm$0.3  &  -4.6$\pm$1.0 &  13.46$\pm$0.01 & -2$\pm$1  &  1.80$\pm$0.03 & 0.02$\pm$0.01  \\ 
(P) $\Delta V_\infty$ mm/s  &4.07  & -5.26 &13.45 & -0.76  &  0.86 & -4.56  \\
(P) $\Delta \theta$ deg  &1  & 1 &2 & 4  &  5 & -   \\
\hline \hline

\end{tabular}}

\caption{\small Earth flyby parameters from \cite{And2008} for spacecraft Galileo (GLL: flybys I and II), NEAR, Cassini, Rosetta and MESSENGER (M'GER).
$V_\infty$ is the average osculating hyperbolic asymptotic speed, $\alpha$ and $\delta$ are the right ascension and declination of the incoming (i) and outgoing (f) osculating asymptotic velocity vectors, and (O) $\Delta V_\infty$ is the  putative ``excess speed"  anomaly deduced by assuming that the speed of light is isotropic in modeling the doppler shifts, as in (\ref{eqn:olddoppler}).   The  observed (O) $\Delta V_\infty$  values are  from \cite{And2008}, and after correcting for atmospheric drag in the case of GLL-II, and thruster burn in the case of Cassini.  (P) $\Delta V_\infty$   is the predicted  ``excess speed", using (\ref{eqn:anomaly}),  after least-squares best-fitting that data using (\ref{eqn:speed}): $\alpha_v$ and $\delta_v$ and $v$ are the right ascension, declination and the 3-space flow speed for each flyby date, which take account of the earth-orbit aberration and  earth inflow effects, and correspond to a galactic flow with $\alpha=4.29$hrs, $\delta=-75.0^\circ$ and $v=486$km/s in the solar system frame of reference.  $\Delta \theta$ is the error, in the best fit, for the aberration determined flow direction, from the nearest flyby flow direction.  In the fitting the MESSENGER data is not used, as the data appears to be anomalous.  }
\label{table:flyby}
\end{table*}

\section{Earth 3-Space Inflow:  Pound and Rebka Experiment\label{sect:PB}}
The numerous EM anisotropy experiments discussed herein demonstrate that a dynamical 3-space exists, and that the speed of the earth wrt this speeds exceeds 1 part in 1000, namely a  large effect.  Not surprisingly this has indeed been detected many times over the last 120 years.  The speed of  nearly  $500$ km/s means that earth based clocks experience a real, so-called, time dilation effect from (\ref{eqn:propertime}) of   approximately 0.12s per day compared to cosmic time.  However clocks may be corrected for this clock dilation effect because their speed $v$ though space, which causes their slowing, is measurable by various experimental methods. This means that the absolute or cosmic time of the universe is measurable.  This very much changes our understanding of time.  However because of the inhomogeneity of the earth 3-space in-flow component the clock slowing effect causes a  differential effect for clocks at different  heights above the earth's surface.  It was this effect that Pound and  Rebka reported in 1960 using the Harvard tower \cite{PB60}. Consider two clocks at heights $h_1$ and $h_2$, with $h=h_2-h_1$, then the frequency differential follows from (\ref{eqn:propertime}),
\begin{eqnarray}
\frac{\Delta f}{f}&=&\sqrt{1-\frac{v^2(h_2)}{c^2}}-\sqrt{1-\frac{v^2(h_1)}{c^2}}  \nonumber \\
&\approx& \frac{v^2(h_1)-v^2(h_2)}{2c^2} +..\nonumber \\
&=& \frac{1}{2c^2}\frac{dv^2(r)}{dr}h+..\nonumber\\
&=& \frac{g(r)h}{c^2}+..\nonumber \\
&=&-\frac{\Delta\Phi}{c^2}+..
\label{PR}\end{eqnarray}
using (\ref{eqn:acceln}) with $\displaystyle{{\bf v.\nabla}{\bf v}=\nabla\left(\frac{v^2}{2}\right) }$ for zero vorticity $\nabla \times {\bf v} = {\bf 0}$, and ignoring any time dependence of the flow, and where finally, $\Delta \Phi$ is the change in the  gravitational potential.  The actual process here is that, say, photons are emitted at the top of the tower with frequency $f$ and reach the bottom detector with the same frequency $f$ - there is no change in the frequency. This follows from (\ref{eqn:period}) but with now $V=0$ giving $T=T^\prime$. However the bottom clock is running slower because the speed of space there is faster, and so this clock determines that the falling photon has a higher frequency, ie. appears blue shifted.  The opposite effect is seen for upward travelling photons, namely an apparent red shift as observed by the top clock.
In practice the Pound-Rebka experiment used motion induced doppler shifts to make these measurements using the M\"{o}ssbauer effect.  The overall conclusion is that Pound and Rebka measured the derivative of $v^2$ wrt to height, whereas herein we have measured that actual speed, but averaged wrt the SC trajectory measurement protocol.  It is important to note that the so-called  ``time dilation" effect is really a ``clock slowing" effect - clocks are simply slowed by their movement through 3-space.  The Gravity Probe A experiment \cite{GPA} also studied the clock slowing effect, though again interpreted differently therein, and again complicated by additional doppler effects.

\section{CMB Direction}
The Cosmic Microwave Background (CMB) velocity  is often confused with the Absolute Motion (AM) velocity or light-speed anisotropy velocity  as determined in the experiments discussed herein. However these are  unrelated and in fact point in very different directions, being almost at 90$^0$ to each other, with the CMB velocity being $369$ km/s in direction $(\alpha=11.2^h, \delta=-7.22^0)$.   The CMB velocity vector was first determined in 1977 by Smoot {\it et al.} \cite{Smoot}.

The CMB velocity is obtained by defining  a frame  of reference in which the thermalised CMB $3^0$K radiation is isotropic, that is by removing the dipole component, and  the CMB velocity is the velocity of the Earth in that frame.    The CMB velocity is a measure of the motion of the solar system relative to  the last scattering  surface (a spherical shell) of the universe  some 13.4Gyrs in the past.  The concept here is that at the time of decoupling of this radiation from matter that matter was on the whole, apart from small observable fluctuations, on average at rest with respect to the 3-space. So the CMB velocity is  not motion with respect to the  {\it local} 3-space now; that is the AM velocity.   Contributions to the AM  velocity would arise from the orbital motion of the solar system within the Milky Way galaxy, which has  a speed of some 250 km/s, and contributions from the motion of the Milky Way within the local cluster, and so on to perhaps super clusters, as well as flows of space associated with gravity in the Milky Way and local galactic cluster etc.    The difference between the CMB velocity and the AM velocity is explained by the spatial flows that are responsible for  gravity at the galactic scales.

\section{Conclusions}
We have shown that the NASA/JPL optical fiber and earth spacecraft flyby data give another independent determination of the velocity of the solar system through a dynamical 3-space. The resulting direction is in remarkable agreement with the direction determined by Miller in 1925/26 using a gas-mode Michelson interferometer. The Miller speed requires a better knowledge of the refractive index of the air atop Mt. Wilson, where Miller performed his experiments, but even using the STP value we obtain reasonable agreement with the NASA/JPL determined speed.  Using an air refractive index of 1.00026 in place of the STP value of 1.00029 would bring the Miller speed into agreement with the NASA data determined speed. As well the NASA/JPL data has permitted the first direct measurement  of the flow of 3-space into the earth, albeit averaged over  spacecraft trajectory during their flybys. This is possible because the inflow component  is radially inward  and so changes direction relative to the other flow components during a flyby, making the flyby doppler shifts sensitive to the inflow speed.

It must be emphasised that the long-standing and repeated determinations of the anisotropy of vacuum EM radiation is not in itself in contradiction with the Special Relativity formalism - rather SR uses a different choice of space and time variables from those used herein, a choice which  by construction mandates that the speed of EM radiation in vacuum be invariant wrt to that choice of coordinates
\cite{CahillMink}. However that means that the SR formalism cannot be used to analyse  EM radiation anisotropy data, and in particular the flyby doppler shift data.

The discovery of absolute motion wrt a dynamical 3-space has profound implications for fundamental physics, particularly for our understanding of gravity and cosmology. It shows that clocks, and all oscillators, whether they be classical or quantum, exhibit a slowing phenomenon, determined by their absolute speed though the dynamical 3-space. This ``clock slowing" has been known as the ``time dilation"  effect - but now receives greater clarity. It shows that there is an absolute or cosmic time, and which can be measured by using any clock in conjunction with an absolute speed detector - many of which have been mentioned herein, and which permits the ``clock slowing" effect to be compensated.  This in turn implies that the universe is a far more coherent and non-locally connected process than previously realised, although a model for this has been proposed \cite{Book}.  It also shows that the now standard discussion of the limitations of simultaneity were really misleading - being based on the special space and time coordinates invoked in the SR formalism, and that simultaneity is a fact of the universe, albeit an astounding one.

As well  successful  absolute motion experiments have always shown wave or turbulence phenomena, and at a significant scale.  This is a new phenomena that is predicted by the dynamical theory of 3-space.  Ongoing development of new experimental techniques to detect and characterise these wave phenomena are reported in \cite{CahillFresnel} .


\begin{thebibliography}{99}


\bibitem{Book} Cahill  R.T. {\it Process Physics: From Information Theory to Quantum Space
       and Matter},  {\it Nova Science Pub.}, New York, 2005.
       
       \bibitem{Review}  Cahill R.T. {\it Dynamical 3-Space: A Review},   in {\it Ether Space-time and Cosmology: New Insights into a Key Physical Medium},   Duffy M. and L\'{e}vy  J., eds.,  {\it Apeiron}, 135-200, 2009.
       
\bibitem{CahillMink}  Cahill R.T. {\it Unravelling Lorentz Covariance and the Spacetime Formalism}, {\it Progress in Physics}, {\bf 4}, 19-24, 2008.



\bibitem{cavities} Braxmaier C. {\it et al.}  {\it Phys. Rev. Lett.}, 88, 010401, 2002;
M\"{u}ller H. {\it et al.} {\it Phys. Rev.} D, 68, 116006-1-17, 2003;  M\"{u}ller H. {\it et al.}  {\it Phys. Rev.}  D67, 056006,2003;
Wolf P. {\it et al.}  {\it Phys. Rev.}  D, 70, 051902-1-4, 2004;  Wolf P. {\it et al.} {\it Phys. Rev. Lett.} , 90, no. 6, 060402, 2003; Lipa J.A., {\it et al.}  {\it Phys. Rev. Lett.}, 90, 060403, 2003.

\bibitem{MMCK} Cahill R.T. and Kitto K. {\it Michelson-Morley Experiments Revisited}, {\it Apeiron}, {\bf 10}(2),104-117, 2003.
 \bibitem{MMC}  Cahill  R.T. {\it The Michelson and Morley 1887 Experiment and the Discovery of Absolute Motion},   {\it Progress in Physics},  {\bf 3}, 25-29, 2005.

\bibitem{MM}  Michelson A.A. and  Morley E.W. {\it Am. J. Sc.} {\bf 34}, 333-345, 1887.
\bibitem{Miller}   Miller D.C. {\it Rev. Mod. Phys.},  {\bf 5}, 203-242, 1933.
\bibitem{Illingworth}   Illingworth K.K. {\it  Phys. Rev.} 3,  692-696, 1927.
\bibitem{Joos}   Joos G. {\it  Ann. d. Physik} [5] 7,  385, 1930.
\bibitem{Jaseja}   Jaseja T.S. {\it et al.} {\it  Phys. Rev.} A 133, 1221, 1964.
\bibitem{Torr}  Torr D.G. and Kolen P. in  {\it Precision Measurements and Fundamental Constants},  Taylor, B.N. and  Phillips, W.D.  eds. {\it  Natl. Bur. Stand. (U.S.), Spec. Pub.}, 617,  675, 1984.
\bibitem{DeWitte}  Cahill R.T. {\it The Roland DeWitte 1991 Experiment}, {\it Progress in Physics}, {\bf 3}, 60-65, 2006.
\bibitem{CahillOFRF} Cahill R.T. {\it  A New Light-Speed Anisotropy Experiment: Absolute Motion and Gravitational Waves Detected},  {\it Progress in Physics}, {\bf 4}, 73-92, 2006,
\bibitem{Munera} Mun\'{e}ra H.A.,  {\it et al.}  in {\it Proceedings of SPIE}, vol 6664, K1- K8, 2007, eds. Roychoudhuri C. {\it et al.} 

\bibitem{CahillOF} Cahill R.T. {\it  Optical-Fiber Gravitational Wave Detector: Dynamical 3-Space Turbulence Detected},  {\it Progress in Physics}, {\bf 4}, 63-68, 2007,

\bibitem{CahillStokesOF}  Cahill R.T.  and Stokes F. {\it  Correlated Detection of sub-mHz Gravitational Waves by Two Optical-Fiber Interferometers}, {\it Progress in Physics}, {\bf 2}, 103-110, 2008.

\bibitem{CahillFresnel}  Cahill R.T.   {\it  Detection of sub-mHz Gravitational Waves using Fresnel Drag Anomaly in Optical Fibers and RF Coaxial Cables}, 2009.



\bibitem{Krisher}    Krisher T.P., Maleki L., Lutes G.F., Primas L.E., Logan R.T., Anderson J.D. and Will C.M. {\it  Test of the Isotropy of the One-Way Speed of Light using Hydrogen-Maser Frequency Standards}, {\it Phys Rev D}, {\bf 42},  731-734, 1990.
\bibitem{CahillFlyby} Cahill R.T. {\it Resolving Spacecraft Earth-Flyby Anomalies with Measured Light Speed Anistropy}, {\it Progress in Physics}, {\bf 4}, 9-15, 2008.
\bibitem{And2008} Anderson J.D., Campbell J.K., Ekelund J.E., Ellis J. and Jordan J.F. {\it  Anomalous Orbital-Energy Changes Observed during Spaceraft Flybys of Earth},  {\it Phys. Rev. Lett.}, {\bf 100}, 091102, 2008.





\bibitem{CahillGW2009}  Cahill, R.T. {\it Quantum Foam, Gravity and Gravitational Waves }, in {\it Relativity, Gravitation, Cosmology: New Developments},  Dvoeglazov V., ed., Nova Science Pub., New York, 2009.

\bibitem{Hicks}  Hicks W.M.  {\it Phil. Mag.}, v.\,3, 9--42, 1902. 

\bibitem{Hertz}  Hertz H.  {\it On the Fundamental Equations of Electro-Magnetics for Bodies in Motion}, {\it Wiedemann's Ann.}  {\bf 41}, 369, 1890;  {\it Electric Waves, Collection of Scientific Papers,}  {\it Dover Pub.}, New  York, 1962.


       
  \bibitem{Schrod} Cahill R.T. {\it  Dynamical  Fractal  3-Space and the Generalised Schr\"{o}dinger  
Equation: Equivalence Principle and  Vorticity Effects},   {\it Progress in Physics},  {\bf 1}, 27-34, 2006.



\bibitem{QC}  Cahill R.T. {\it A Quantum Cosmology: No Dark Matter, Dark Energy nor Accelerating Universe}, arXiv:0709.2909, 2007.
\bibitem{Paradigm}  Cahill R.T. {\it Unravelling the Dark Matter - Dark Energy Paradigm}, arXiv:0901.4140, 2009.


\bibitem{CahillSuper2009}  Cahill, R.T. {\it The Dynamical Velocity Superposition Effect in the Quantum-Foam Theory of Gravity}, in {\it Relativity, Gravitation, Cosmology: New Developments},  Dvoeglazov V., ed., Nova Science Pub., New York, 2009.


\bibitem{PB60}  Pound R.V. and Rebka Jr. G.A. {\it Phys. Rev. Lett.}, {\bf 4}(7), 337-341, 1960.

\bibitem{RingLaser}  Schreiber K.U., Velikoseltsev A., Rothacher M., KL\"{u}gel T., Stedman G.E. and Wilsthire D.L. {\it Direct Measurement of Diurnal Polar Motion by Ring Laser Gyroscopes}, {\it J. Geophys. Res.} {\bf 109},  B06405, doi:10.1029/2003JB002803. 

\bibitem{Smoot}  Smoot G.F., Gorenstein M.V. and Muller R.A.  {\it Phys. Rev. Lett.}, {\bf 39}(14), 898, 1977.


  \bibitem{Flow} Cahill R.T. {\it  3-Space Inflow Theory of Gravity: Boreholes, Blackholes and the Fine Structure Constant},   {\it Progress in Physics},  {\bf 2}, 9-16, 2006.
  
  \bibitem{CahillQFoamFlow} Cahill R.T. {\it Dark Matter as a Quantum Foam In-Flow Effect}, {\it Trends in Dark Matter Research}, ed. J. Val Blain, Nova Science Pub.,  95-140, NY, 2005. 
  
\bibitem{GPA} Vessot R.F.C. {\it et al.} {\it Test of Relativistic Gravitation with a Space-Borne Hydrogen Maser}, {\it Rev. Mod. Phys.} {\bf 45}, 2081, 1980.

\end{thebibliography}
\end{document}